\documentclass[useAMS,usenatbib,onecolumn]{mn2e}
\usepackage{deluxetable,amssymb,rotating}

\title[Apsidal motions of 90 eccentric EBs in the SMC]{Apsidal motions of 90 eccentric binary systems in the Small Magellanic Cloud}

\author[Kyeongsoo Hong et al.]
       {Kyeongsoo Hong$^{1}$\thanks{E-mail: kshong@kasi.re.kr}, Jae Woo Lee$^{1,2}$, Seung-Lee Kim$^{1,2}$, Jae-Rim Koo$^1$  
       \newauthor        and Chung-UK Lee$^{1,2}$ \\
        $^1$Korea Astronomy and Space Science Institute, 776 Daedeokdae-ro, Yuseong-gu, Daejon 34055, Korea\\
        $^2$Astronomy and Space Science Major, Korea University of Science and Technology, Daejeon 34113, Korea}
\begin{document}

\date{Accepted 201? ---------. Received 201? ---------; in original form 201? }

\pagerange{\pageref{firstpage}--\pageref{lastpage}} \pubyear{2016}

\maketitle

\label{firstpage}

\begin{abstract}
We examined light curves of 1138 stars brighter than 18.0 mag in the $I$ band and less than a mean magnitude error of 0.1 mag in the $V$ band from the OGLE-III eclipsing binary catalogue, and found 90 new binary systems exhibiting apsidal motion. In this study, the samples of apsidal motion stars in the SMC were increased by a factor of about 3 than previously known. In order to determine the period of the apsidal motion for the binaries, we analysed in detail both the light curves and eclipse timings using the MACHO and OGLE photometric database. For the eclipse timing diagrams of the systems, new times of minimum light were derived from the full light curve combined at intervals of one year from the survey data. The new 90 binaries have apsidal motion periods in the range of 12$-$897 years. An additional short-term oscillation was detected in four systems (OGLE-SMC-ECL-1634, 1947, 3035, and 4946), which most likely arises from the existence of a third body orbiting each eclipsing binary. Since the systems presented here are based on homogeneous data and have been analysed in the same way, they are suitable for further statistical analysis.
\end{abstract}


\section{Introduction}

The investigation of apsidal motions in eccentric eclipsing binaries (EBs) is very important because it provides valuable information on the internal structure of the stars. Apsidal motion is caused by the fact that binary components are not point masses, and its magnitude strongly depends on how centrally condensed the stars are. The period of the apsidal motion can be determined by the detailed analyses of both light curves and eclipse timings. Detached, double-lined EBs with an apsidal motion provide an accurate and direct determination of fundamental stellar properties such as mass, radius, and luminosity (Hilditch 2001). We can compute the internal structure constants ($k_2$) from the apsidal periods and absolute dimensions of EBs. These values then allow us to test stellar structure and evolution models and to calculate the distance to the systems (Torres et al. 2010).

The study of massive and metal-deficient stars in the Magellanic Clouds (MCs) can test theoretical models for different abundances, and the large samples of the extragalactic binaries provides clues about the structure and evolution of massive stars with low metallicity (Zasche et al. 2014, Ribas 2004). In particular, the binary systems exhibiting apsidal motions provide useful information on the interior structures of the massive stars. Recent large surveys such as MAssive Compact Halo Objects (MACHO; Faccioli et al. 2007) and Optical Gravitational Lensing Experiment (OGLE; Graczyk et al. 2011, Pawlak et al. 2013) have discovered tens of thousands of new EBs in the MCs, many of which have been found to be in eccentric orbits. Faccioli et al. (2007) presented 4634 and 1509 EBs in the Large Magellanic Cloud (LMC) and the Small Magellanic Cloud (SMC), respectively, from the MACHO survey. Wyrzykowski et al. (2003, 2004) published $V$ and $I$ photometry for 2580 and 1351 EBs in the LMC and the SMC, respectively, from the OGLE-II survey. The EB catalogues from the OGLE-III survey identified 26121 and 6138 binary systems for the LMC (Graczyk et al. 2011) and SMC (Pawlak et al. 2013), respectively. The large samples of about 32000 EBs from the OGLE-III can, therefore, provide us with many candidates showing apsidal motion.

The apsidal motion binaries have increased in number in recent times (Michalska \& Pigulski 2005, Michalska 2007, Zasche \& Wolf 2013, Hong et al. 2014, and Zasche et al. 2015 for the LMC, and Graczyk 2003, North et al. 2010, Zasche et al. 2014 and Hong et al. (2015) for the SMC). However periods of apsidal motion have only been estimated for 21 and 42 EBs in the LMC and the SMC, respectively. The numbers of apsidal binaries in the MCs are still limited for statistical studies. This paper is the third contribution in a series of apsidal motion studies for binary systems in the MCs (Hong et al. 2014, 2015).

The main aim of this paper is to provide the orbital parameters and accurate apsidal motion elements of 90 eccentric binary stars in the SMC using homogeneous data and consistent analysis methods. Section 2 presents the selection method of objects with the large photometric survey data of MACHO and OGLE. In Section 3, we describe the analysis method of light curves. The apsidal motion elements are calculated from the light curves and eclipse timings in Section 4. Finally, Section 5 presents a summary and discussion of this work.

\section{Selection of objects}

During the past 20 years, gravitational lensing surveys have obtained photometric observations for thousands of EBs in the SMC. In this paper, we used data from MACHO (1992-1999; Faccioli et al. 2007), OGLE-II (1997-2000; Wyrzykowski et al. 2004) and OGLE-III (2001-2009; Pawlak et al. 2013). The MACHO observations were obtained in blue ($V_{M}$; $440-590$ nm) and red ($R_{M}$; $590-780$ nm) bands, corresponding roughly to the Johnson $V$ and Cousins $R$, respectively. The MACHO survey used the 1.27-m Great Melbourne Telescope at Mount Stromlo in Australia. The OGLE-II and OGLE-III observations covered 2.4 and 14 square degrees, respectively, in the SMC using the 1.3-m Warsaw Telescope in the $V$ and $I$ bands (Udalski et al. 1997, Szyma\'nski 2005, Udalski et al. 2008). The OGLE-III EB database given by Pawlak et al. (2013) was the primary source of the main selection process. The OGLE data based on Difference Image Analysis (DIA) showed considerably smaller scatter than the MACHO data.

First of all, we extracted 1138 binaries among the 6138 EBs in the SMC, with the selection criterion that the binaries are brighter than 18.0 mag in the $I$ band and have magnitude errors within 0.1 mag in the $V$ band for the spectroscopic observations using 8-m class telescopes (cf. North et al. 2010). In order to find the apsidal motion binaries, we examined the $I$ band light curves of these systems obtained from OGLE-III observations using the iteration method applied in the papers of Kang et al. (2012) and Hong et al. (2014, 2015). The iteration method is a modification of the Wilson \& Devinney (WD) differential correction code (Wilson \& Devinney 1971, hereafter WD) used to treat large numbers of light curves of EBs.

For this procedure, the initial surface temperatures of the selected 1138 EBs were assumed to be 15000 K. In the first step, we adjusted the following parameters: the temperature of the secondary component ($T_{2}$), the surface potentials ($\Omega_{1,2}$), the orbital inclination ($i$), the eccentricity ($e$), the longitude of periastron ($\omega$), the epoch ($T_{0}$), the orbital period ($P$) and the luminosity of the primary star ($L_{1}$). After the analyses, we excluded the binary systems in circular orbits of $e=0$. Then we included the rate of periastron advance ($\dot{\omega}=d\omega / dt$) as a free parameter.
The WD runs were repeated until the correction of each parameter became smaller than its standard deviation.
We determined preliminary apsidal periods for the systems with $\dot{\omega}$ above 0.0 through the method used in Sections 3 and 4. Among these, dozens of EBs have uncertainties larger than their apsidal periods and orbital eccentricities because of the lack of and large scatter of observations. Therefore, we do not included these binaries in subsequent analyses.
Finally, we found 90 binary systems exhibiting apsidal motion, excepting to the 36 EBs studied previously by Zasche et al. (2014) and Hong et al. (2015). The basic information for these stars is listed in Table 1, in which the coordinates, $I$, $V$ and $V-I$ are taken from the OGLE-III of Pawlak et al. (2013). The colour-magnitude diagram of the selected 90 binaries in the SMC is displayed with the well-studied eclipsing binaries with the apsidal motions in Figure 1.

\section{Light curve synthesis}

For photometric solutions, $V_{M}R_{M}VI$ light curves of the 90 EBs were taken from both the MACHO and OGLE-III survey data obtained across 1992-1999 and 2002-2009, respectively. The light curves were analysed in a manner identical to the papers of Hong et al (2014, 2015) by using the 2005 version of the WD code. The initial mass ratios for all selected systems were set to be $q=M_2/M_1=1.0$, because it is difficult to determine a reliable $q$ value from the light-curve analyses of detached EBs (Wyithe \& Wilson 2001, Terrell \& Wilson 2005). We computed the new mass ratio of the 90 EBs from the preliminary photometric solutions and the relation between the mass ratio and the luminosity applied in the papers by Graczyk (2003) and Zasche et al. (2015). In order to obtain the photometric mass ratio, we used the approach as presented below.

\begin{flushleft}
\begin{verse}
1. At the beginning, all light curves for each EB were analysed using $q=1.0$, which resulted in a temperature ratio, and fractional radii ($r_{1}$ and $r_{2}$).
\end{verse}
\begin{verse}
2. The $q-L$ relation is given by ${q} = 10^{({\rm log} L_{2}-{\rm log} L_{1})/3.664}$. For this approach, the luminosities of the selected systems were required. In order to estimate the absolute dimensions for each binary component, we derived the semi-major-axis (a) from the following equation (Graczyk 2003, Hong et al. 2014):
\begin{equation}
{a} = \left(\frac{Luminosity~of~the~ binary~ system}{4\pi\sigma(r_{1}^{2}T_{1}^{4}+r_{2}^{2}T_{2}^{4})}\right)^{1/2},
\end{equation}
where, $L_{system}$ is the total luminosity of a binary system and can be derived by the bolometric magnitude ($M_{bol,system}$) of the system. The fractional radii ($r_{1}$ and $r_{2}$) and temperature ratio ($T_{2}/T_{1}$) of the primary and secondary components were obtained from the light curve solutions. The temperatures of the systems were estimated from the intrinsic colour index $(V-I)_{0}=(V-I)_{obs}-E(V-I)$ using the relation between the colour index and temperature by Worthey \& Lee (2011). The absolute dimensions were determined with the semi-major-axis and assuming a mean distance modulus (DM) of the SMC of $V-M_{V}=19.11$ (North et al. 2010) and that the stars follow the mass-luminosity relation ($M-L$) for the SMC (log $L=3.664$ log $M+0.380$, Graczyk 2003). If the EBs in the SMC are slightly evolved stars, the uncertainties of their absolute dimensions  are increased. From this, the satisfied colour excess $E(V-I)$ in the assumptions was obtained. The method allows the obtainment of an independent reddening and a reasonable $(V-I)_0$ intrinsic colour in the direction of the binary system (see e.q. Graczyk 2003). Note the mean distance modulus of the SMC has systematic uncertainties by the line of sight depth ($4.90 \pm 1.23$ kpc) for the SMC central bar region by Subramanian \& Subramaniam (2009). The new photometric mass ratio was then estimated using the relation between the mass ratios and luminosities of both components.
\end{verse}

\end{flushleft}

These processes were repeated three to five times for each EB until all the conditions (light curve solutions, $M-L$, $DM$ and the relation between mass ratios and luminosities) had been satisfied. In each WD run, the initial values of the parameters were adopted from the photometric solutions with the new mass ratio derived from the previous stage. The logarithmic bolometric ($X_{1,2}$) and monochromatic ($x_{1,2}$) limb-darkening coefficients were interpolated from the values in van Hamme (1993). The gravity darkening exponents and the bolometric albedos were fixed at the standard values of g1 = g2 = 1.0 and A1 = A2 = 1.0 for stars with radiative envelopes from their temperatures. We used mode 2 (detached system) and the standard deviation ($\chi^{2}$) between observations and theoretical modes were calculated to measure the goodness of each model. The final model was the one with the lowest $\chi^{2}$ value.

We searched for the possible presence of a third light ($l_3$) in the light curves of all selected systems using the differences of $\chi^{2}$ between the third light and the non-third light models. The final results are listed in Table 2 and Table C1 (only light ratios are listed), where the uncertainty of each parameter was calculated by means of the Differential Corrections subroutine in the WD code. Note that the parameter errors in the WD code are estimated from the covariance matrix evaluated around a best-fitting model. It should also be noted that the error estimates are unrealistically small due to the strong correlations between the relatively numerous parameters. The third lights could come from a tertiary star physically bound to or only optically related with each EB. The $V_{M}R_{M}VI$ observations and the model light curves for the 90 EBs are plotted in Figure 2 and Appendix A as normalised lights versus orbital phases. As one can see, the eclipses of the primary and secondary in the MACHO data have moved to the phase decreasing or increasing direction in the OGLE data due to apsidal motion. The resulting absolute dimensions with the colour excess are listed in Table D1.

\section{Apsidal motion analysis}

The periods of the apsidal motions were obtained from the detailed analyses of both light curves and eclipse timings. The apsidal periods ($U$) from the former were calculated from the rate of periastron advance ($\dot{\omega} = 2 \pi / U$) in $V_{M}R_{M}VI$ light curve solutions. For the eclipse timing analysis, the times of minimum light were derived from the full light curve formed from the observation points at an interval of 1 year because the surveys provided only two to five data points per night for an observing field. 
Each light curve from the MACHO, OGLE-II, and OGLE-III observations was analysed using the iteration method with the photometric solutions in Table 2. In total, 2224 eclipse times were obtained from the light curves, and the eclipse timing method is the same as that described by Hong et al. (2014, 2015). The apsidal motion can be described as having five apsidal elements: zero epoch ($T_{0}$), sidereal period ($P_{s}$), anomalistic period ($P_{a}$), rate of periastron advance in degree per cycle ($\dot{\omega}$), and eccentricity ($e$). The apsidal motion elements of the selected systems were derived from the ephemeris-curve equation presented by Gim\'enez \& Bastero (1995). 
All individual minimum times are provided in Table 3, and the O-C eclipse timing diagrams with respect to the apsidal motion equation are plotted in Figure 3 and Appendix B. The resulting elements of apsidal motions and their errors are given in Table 4.

As shown in the eclipse timing diagrams, the apsidal motions of OGLE-SMC-ECL-1634, 1947, 3035, and 4946 could not describe their minimum times satisfactorily and the residuals in the lower panel indicate the existence of an additional short-term oscillation.
Thus, we fitted the timing residuals to a sine curve, $C=K{\rm sin}(\Omega E+ \Omega_0)$. The Levenberg-Marquardt method (Press et al. 1992) was applied to solve the three parameters of the ephemeris given in Table 5, together with related quantities. The results are shown in the middle panels of Figure 4. The top panels indicate the combination effect of the apsidal motion and the oscillation for each system, and the bottom panels represent the residuals from the ephemeris. As can be seen in this figure, the entire collection of timings can be better fitted by a beat effect of the two effects. 

The sinusoidal variation of the eclipse timings could be produced by one of two causes: a light-travel-time effect due to a third body (Irwin 1952, 1959) or a magnetic activity cycle of one or both of the component stars (Applegate 1992, Lanza et al. 1998). However, the magnetic mechanism never displays a pattern of alternating period decreases and increases for systems with the spectral type earlier than about F5. This implies that the oscillations in the eclipsing timings may originate from the presence of a third object physically bound to the close binaries. Furthermore, our light-curve synthesis detected the significant contribution of the third light in these systems. These systems may be triple stars. In Table 5, $P_3$ and $a_{12}$sin$i_3$ are the cycle length and semi-major axis of the eclipsing pair around the mass centre of the triple systems, respectively, and f($M_3$) is the mass function of the third body.

\section{Summary and discussion}

We found 90 previously unknown EBs showing apsidal motions from 6138 binary systems brighter than $I=18$ mag and more precise than 0.1 mag in OGLE-III survey data of the SMC. Their photometric solutions and elements of the apsidal motions were derived for the first time. An additional short-term oscillation was detected for five systems (OGLE-SMC-ECL-1634, 1947, 3035, 4946, and 5382) which makes a considerable contribution to the third light in our light curve synthesis. The period modulations may be interpreted as a light-travel-time effect due to a circumbinary object gravitationally bound to the systems. Including our results, the number of systems with apsidal motions in the SMC has significantly increased than previously known. All of the selected EBs were modelled by using the homogeneous data and the same analysis method and they can be used for a statistical study of the stars in the SMC, as well as for testing the theoretical models of stellar structure and evolution in the low-metallicity environment. The histogram of the apsidal periods of the 126 SMC EBs is presented in Figure 5, together with those of 88 detached binaries of our Galaxy in Table E1 for comparison. As shown in the figure, all EBs in the SMC have apsidal periods in the range of 1.0 $<$ log $U$ $<$ 3.0 because of the selection effect from their short-term observations with a time span of about 20 years.

From the OGLE LMC data, Mazeh et al. (2008) suggested that the apsidal motion periods should depend mostly on the binary period. In order to examine this possibility in the SMC, we plotted the apsidal periods versus the orbital periods with the known 100 EBs in Figure 6. Their physical parameters are listed in Table E1. In Figure 6, we have plotted the predicted values of the apsidal motion periods against the orbital periods for the eccentricities of 0.001, 0.3 and 0.5, having two masses of 1.4 and 1.8 $M_{\odot}$. The apsidal motion rate ($\dot\omega = \dot\omega_{\rm N} + \dot\omega_{\rm R}$) for the six cases was calculated using the equations (3) and (11) given in the paper of Gim\'enez (1985). The mass ratios were assumed to be $q=m_2/m_1=1.0$ for these calculations. We can see that the apsidal period indicates a linear dependence on the binary orbital period. The massive binaries in the SMC and LMC are placed between the intermediate and high mass stars. However, there are slight trend differences between the EBs of SMC and our Galaxy. It is possible that the differences may come from the separate age and metallicity between them. In order to understand the difference, we need to determine the accurate physical properties for the SMC EBs, as well as to find a large number of long-period EBs ($P$ $>$ 6 d) that exhibit apsidal motions.

Figure 7 shows the eccentricities of the 126 EBs in the SMC as a function of their orbital periods. The dashed curve was obtained with the equation $f(P)=E-A\times {\rm exp}^{-(P\times B)^C}$ from Mazeh et al. (2008), where $P$ denotes the orbital period and the other parameters are $E=0.98$, $A=3.25$, $B=6.3$ and $C=0.23$. In this figure, we can see that all binaries exhibiting apsidal motions in the SMC have their eccentricities below the curve, while there are no eccentric EBs with orbital periods shorter than about 1 d. This implies that short-period EBs are in circular orbits, and the cause of their absence may be the result of tidal circularisation, which is substantially weaker in longer period binaries. The 126 EBs with a binary period less than about 7 d in the SMC also support the suggestion of Mazeh et al. (2008). Because no spectroscopy has been conducted for almost all of the EBs with apsidal motion in the SMC, spectroscopic observations are required for more detailed statistical studies of the various physical parameters.

\section{Acknowledgments}
We thank the anonymous reviewers for very useful comments, which helped us to improve the manuscript. We are grateful to the MACHO and OGLE teams for making their excellent photometric data base publicly available. This work was supported by KASI (Korea Astronomy and Space Science Institute) grant 2016-1-832-01.

\clearpage 
\begin{figure*}
\begin{center}
\begin{tabular}{c}
\includegraphics{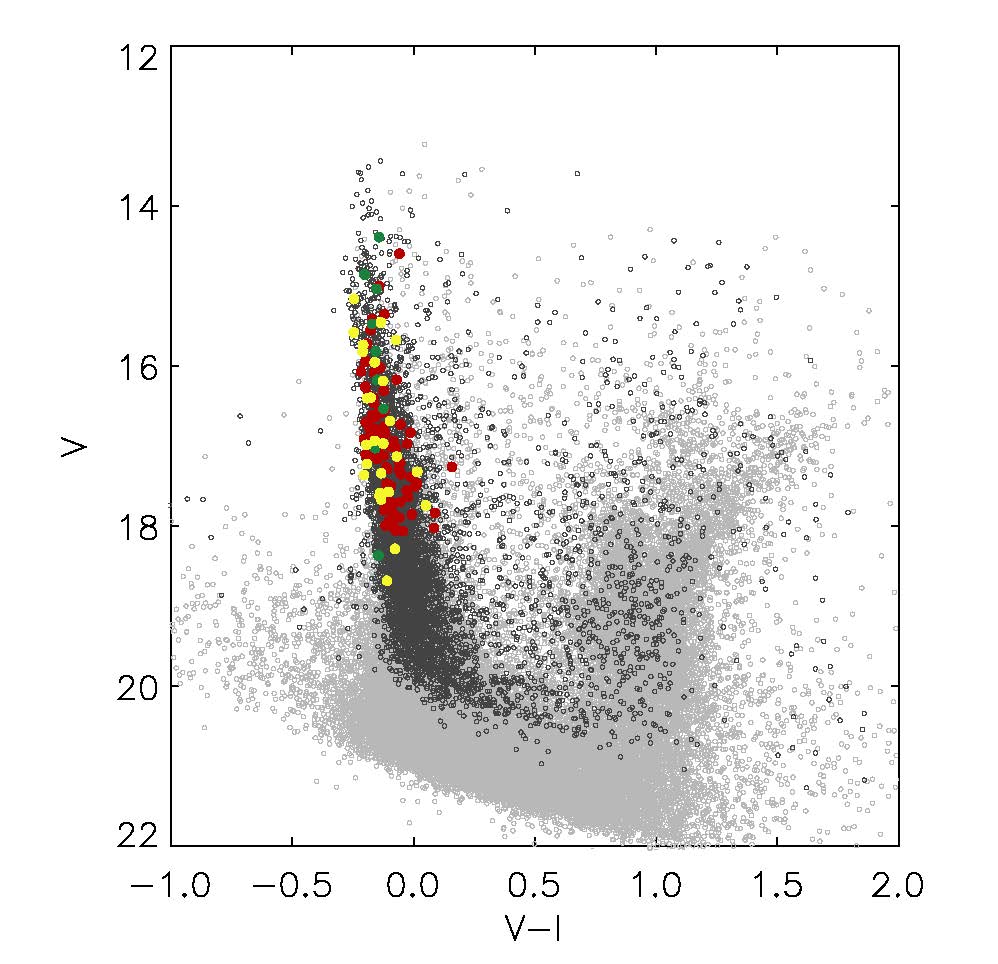}  
\end{tabular}
\caption{The colour-magnitude diagram for $\sim$56000 stars (grey circles) from the OGLE-III 100.1 field in the SMC.
The black circles represent the 6138 EBs given by Pawlak et al. (2013). 
The red circles represent the 90 EBs found in this paper.
The yellow and green circles display the 27 EBs by Hong et al. (2015) and 18 EBs by Zasche et al. (2014), wherein nine systems are overlapped each other. }
\end{center}
\end{figure*}

\clearpage
\begin{figure}
\centering 
\begin{center}
\begin{tabular}{ccc}
\includegraphics[width=0.3\columnwidth,height=6.5cm]{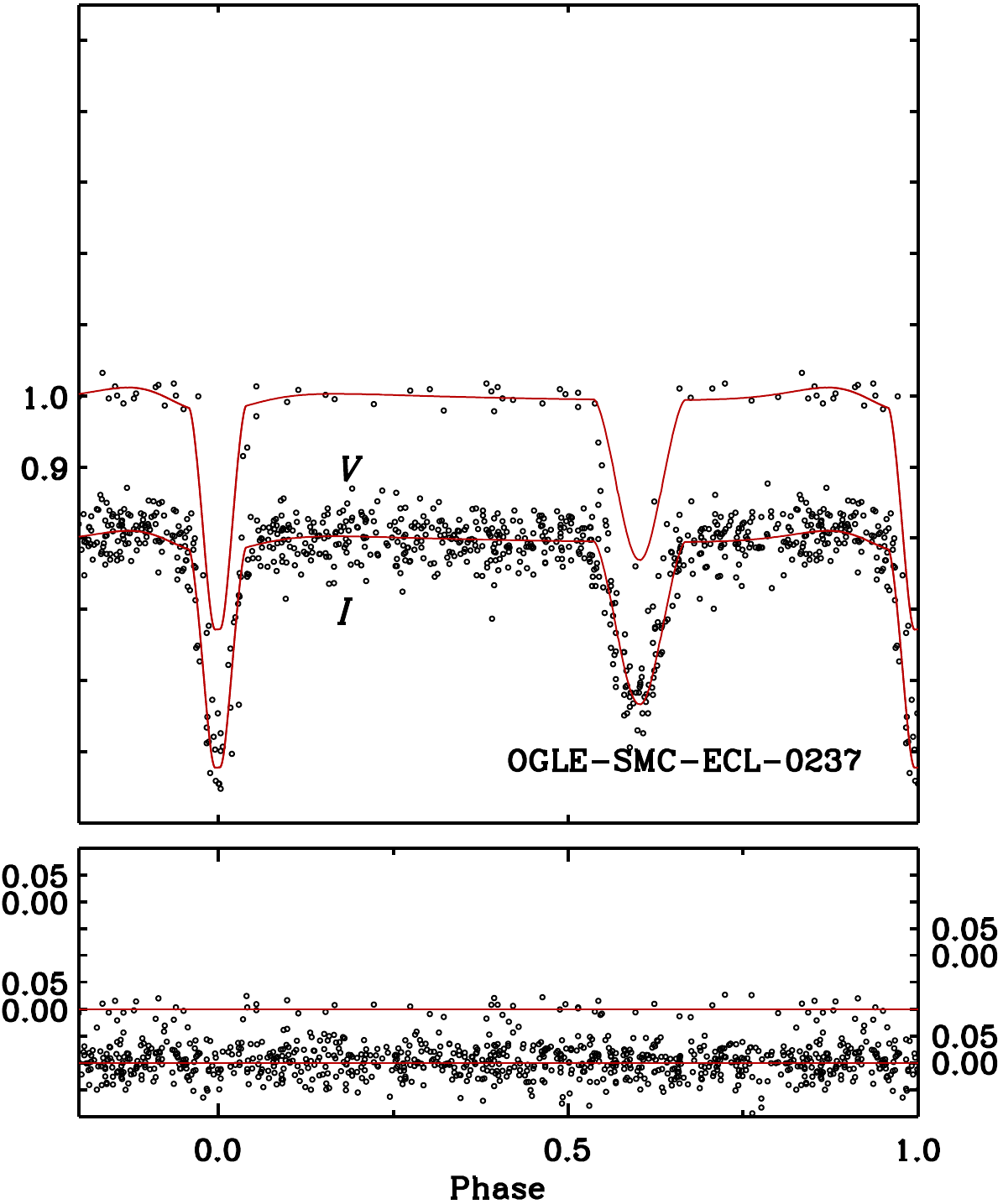} & \includegraphics[width=0.3\columnwidth,height=6.5cm]{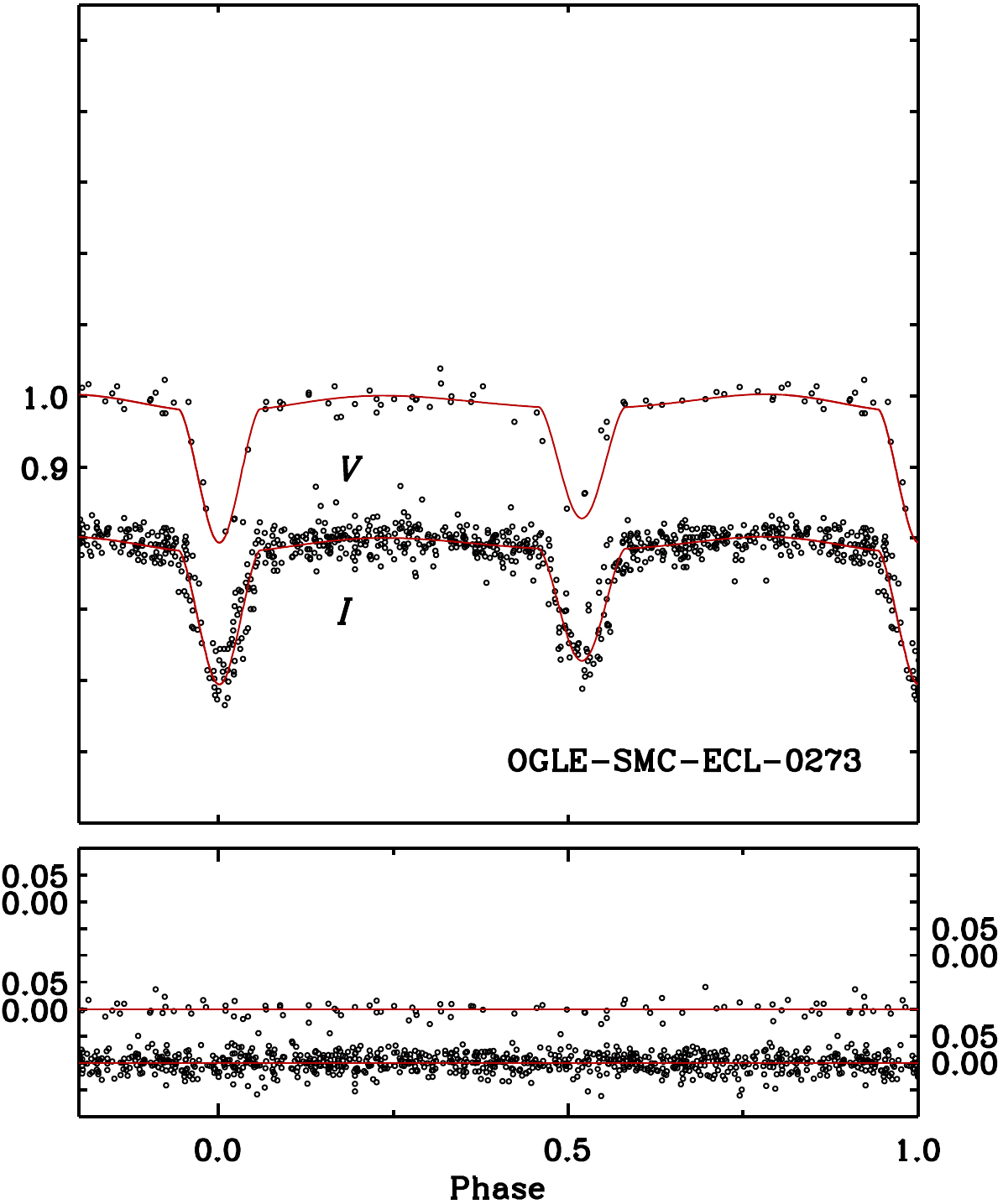} & \includegraphics[width=0.3\columnwidth,height=6.5cm]{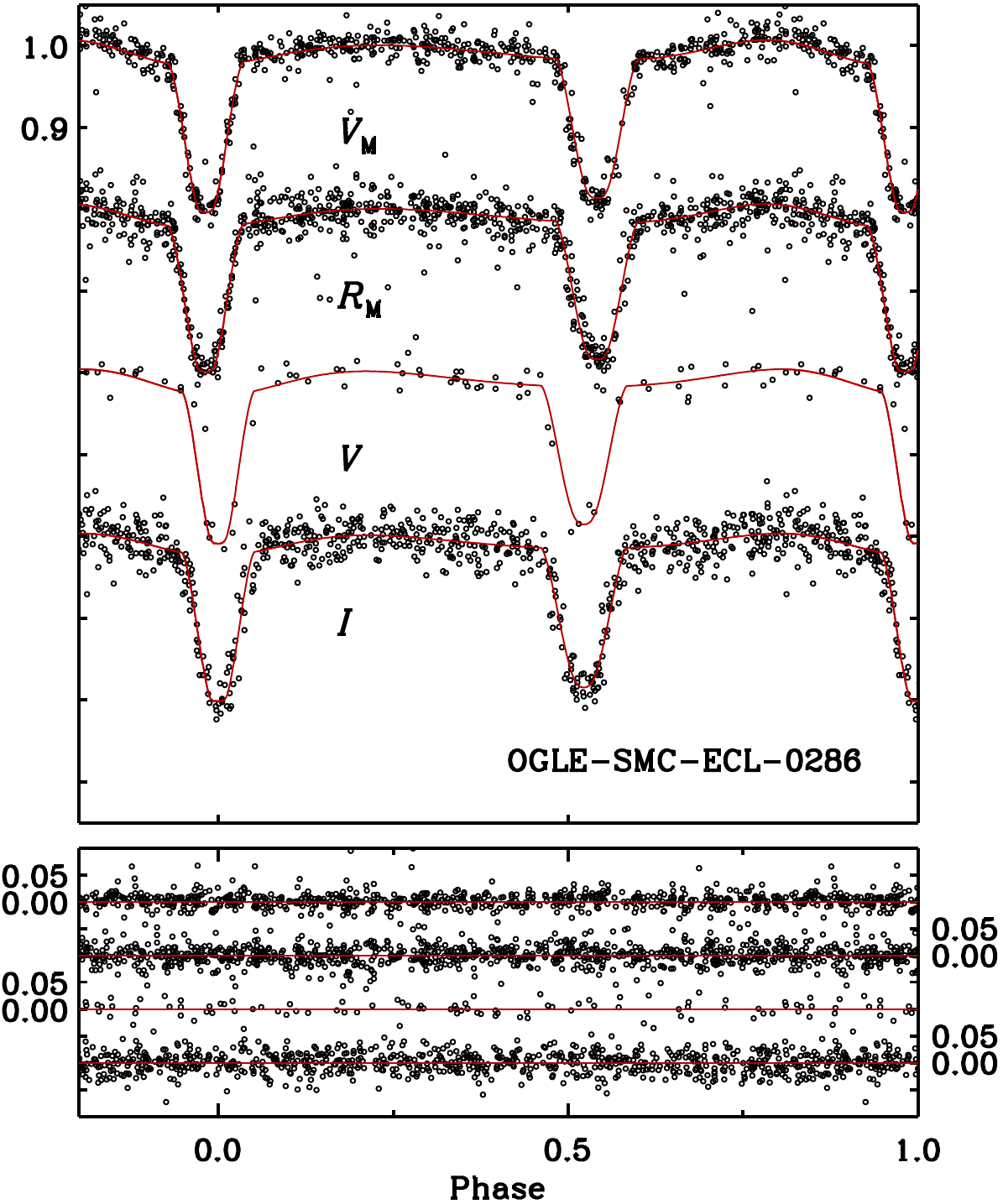}  \\
\includegraphics[width=0.3\columnwidth,height=6.5cm]{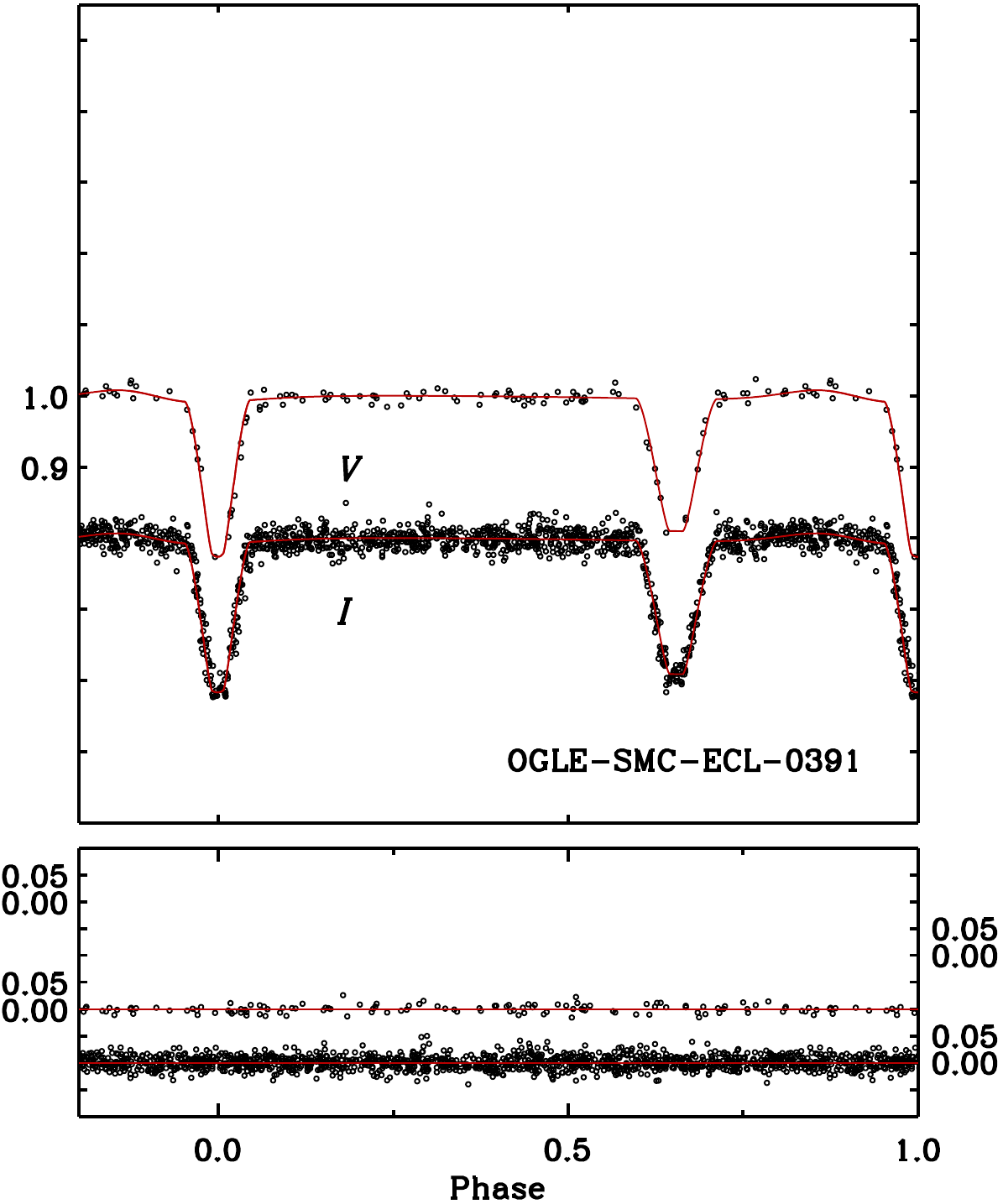} & \includegraphics[width=0.3\columnwidth,height=6.5cm]{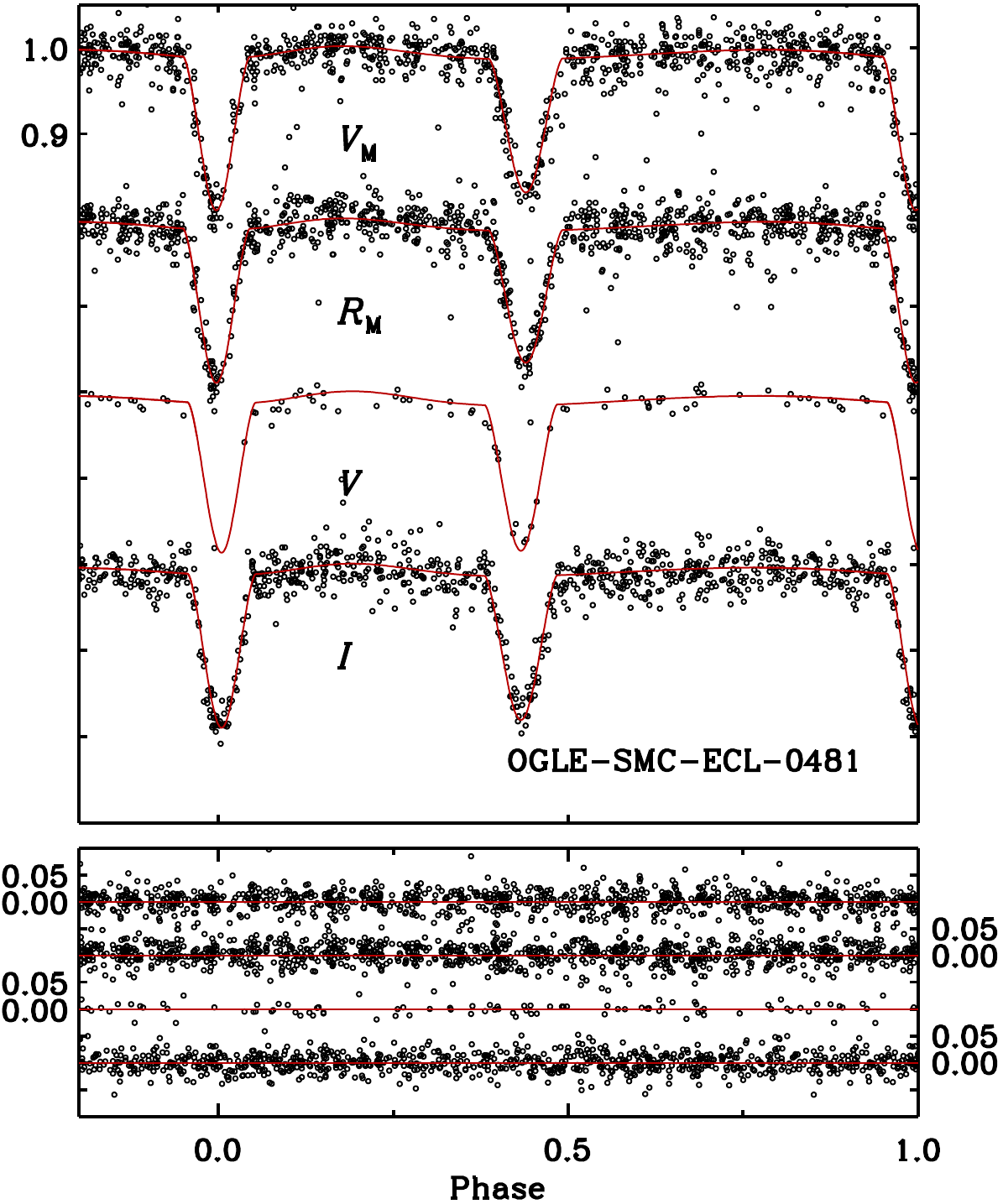} & \includegraphics[width=0.3\columnwidth,height=6.5cm]{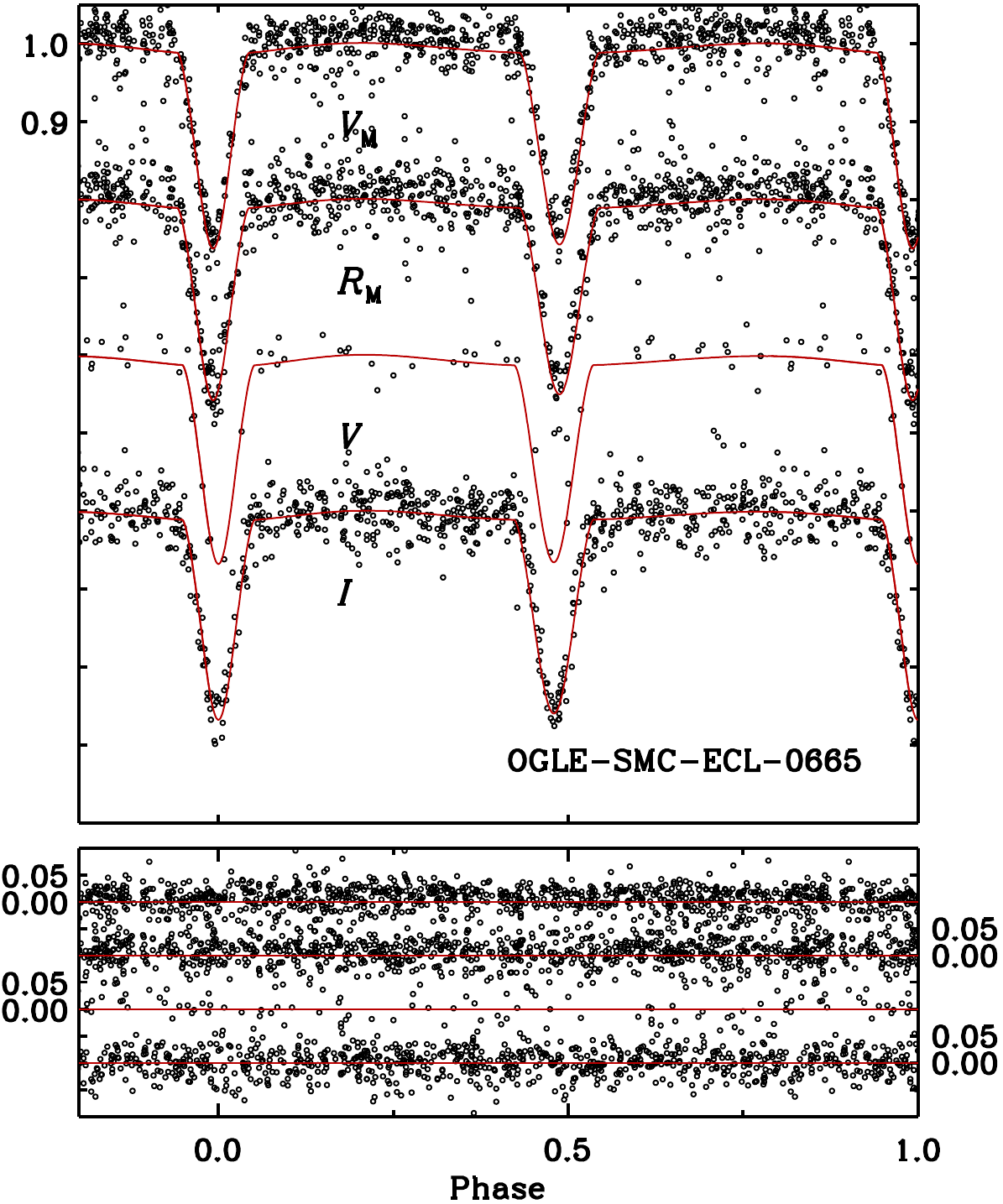}  \\
\includegraphics[width=0.3\columnwidth,height=6.5cm]{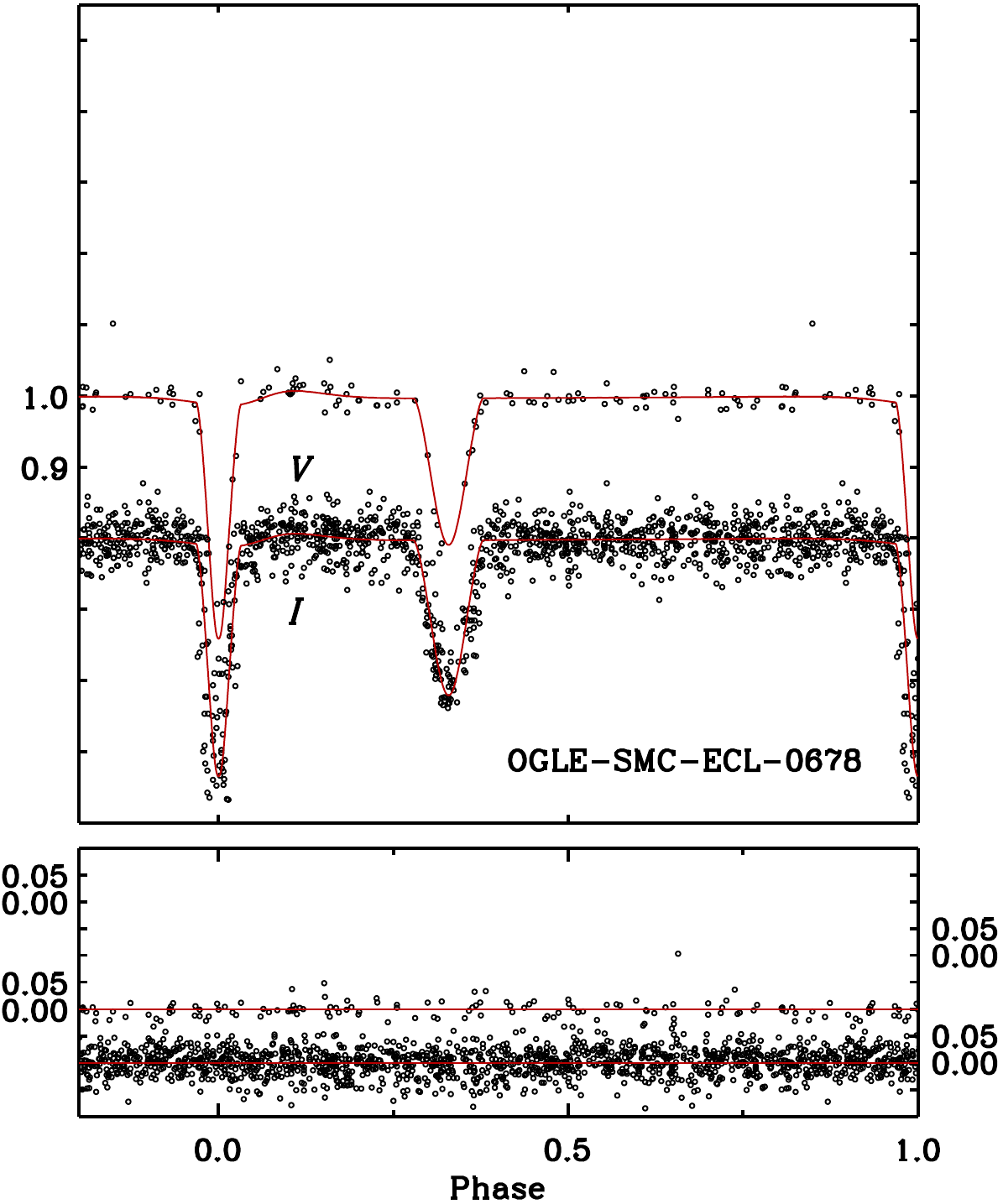} & \includegraphics[width=0.3\columnwidth,height=6.5cm]{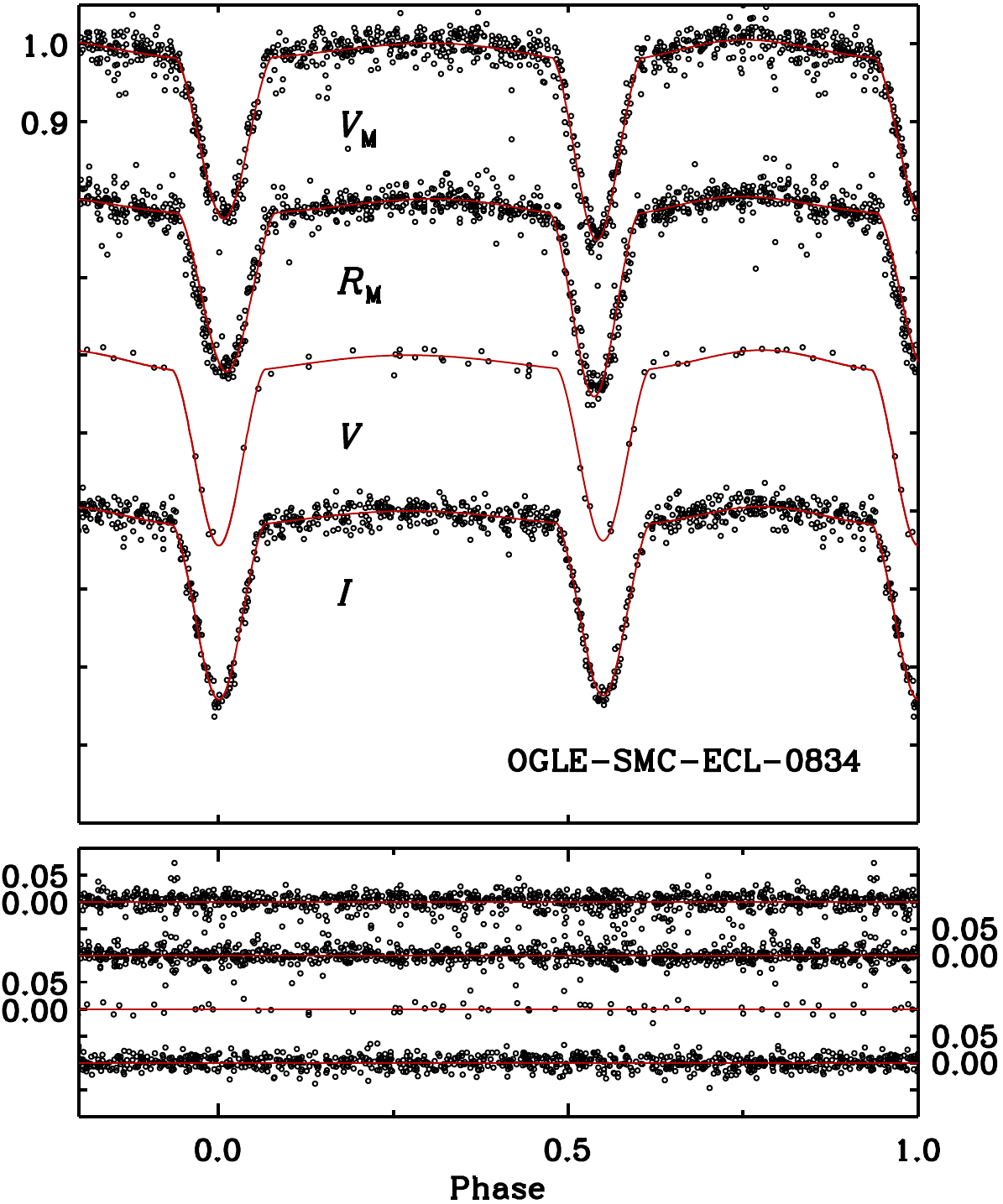} & \includegraphics[width=0.3\columnwidth,height=6.5cm]{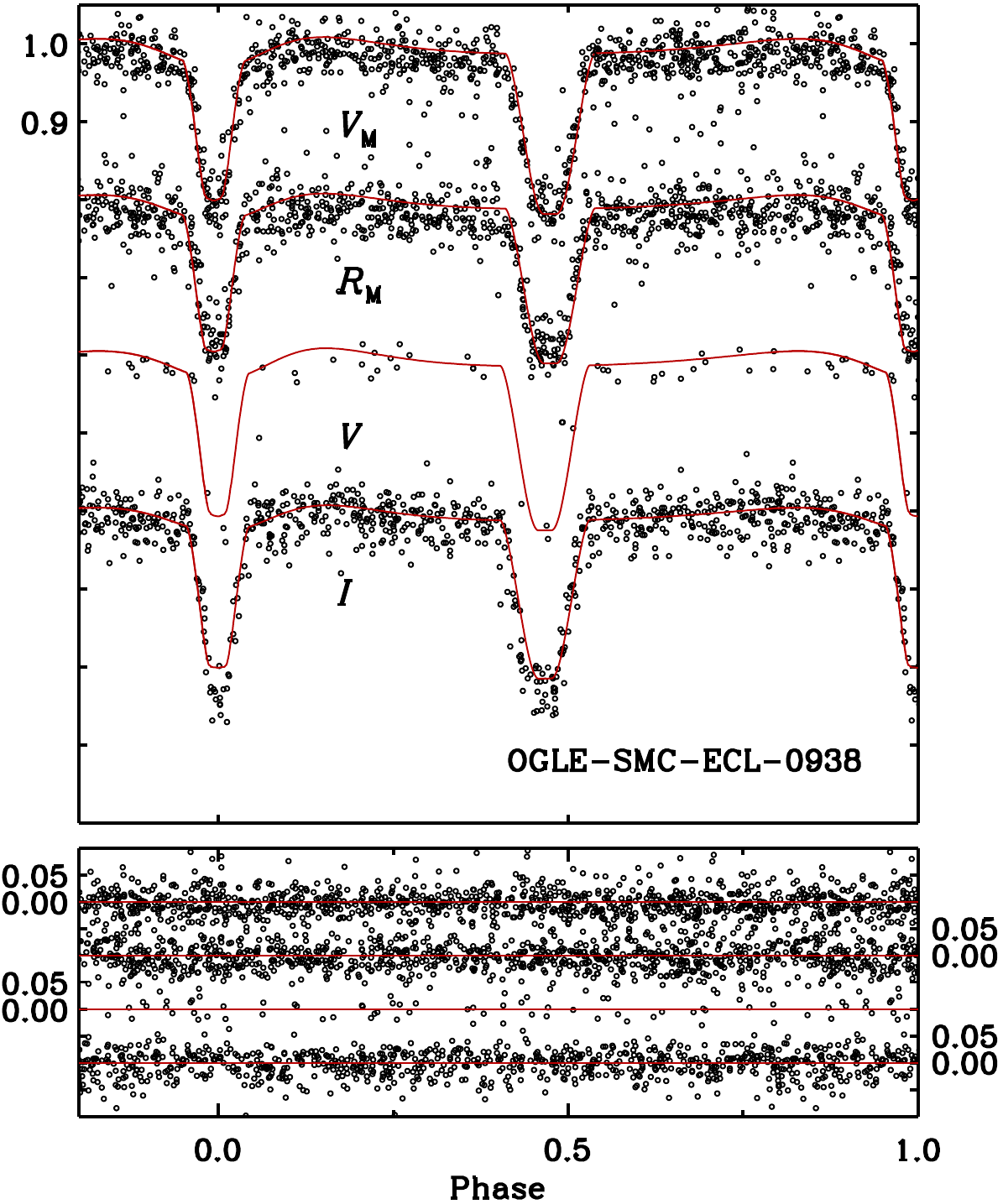}  \\
\end{tabular}
\caption{$V_{M}R_{M}VI$ light curves of 9 binary systems. The upper panels display light curves using the MACHO and OGLE-III data, wherein the open circles are the observations and the solid lines are the theoretical light curves. The lower panels are the light residuals between observations and theoretical models. The open circles are the observations.}
\end{center}
\end{figure}

\clearpage

\begin{figure}
\begin{center}
\begin{tabular}{ccc}
\includegraphics[width=0.3\columnwidth,height=4.5cm]{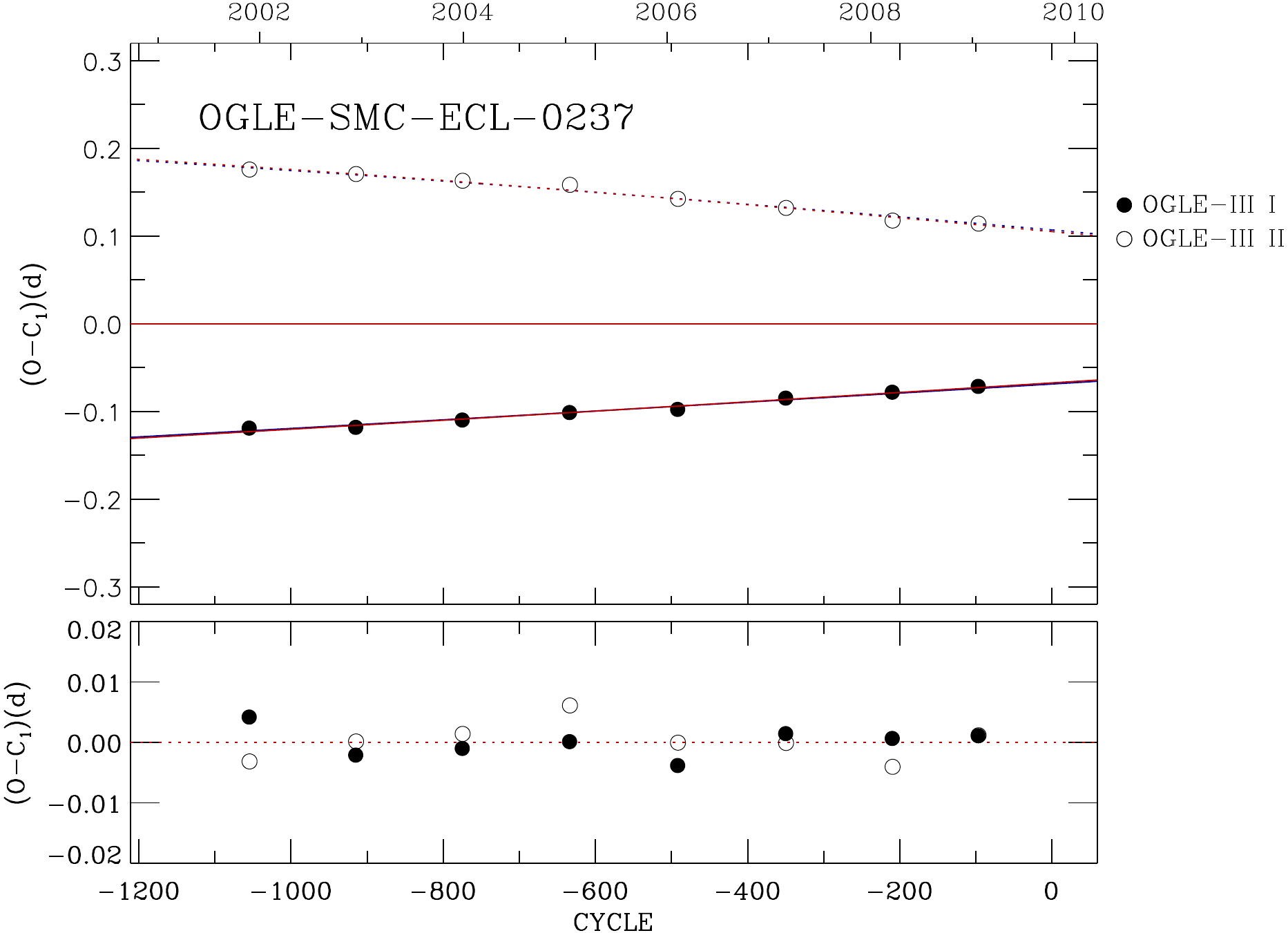} & \includegraphics[width=0.3\columnwidth,height=4.5cm]{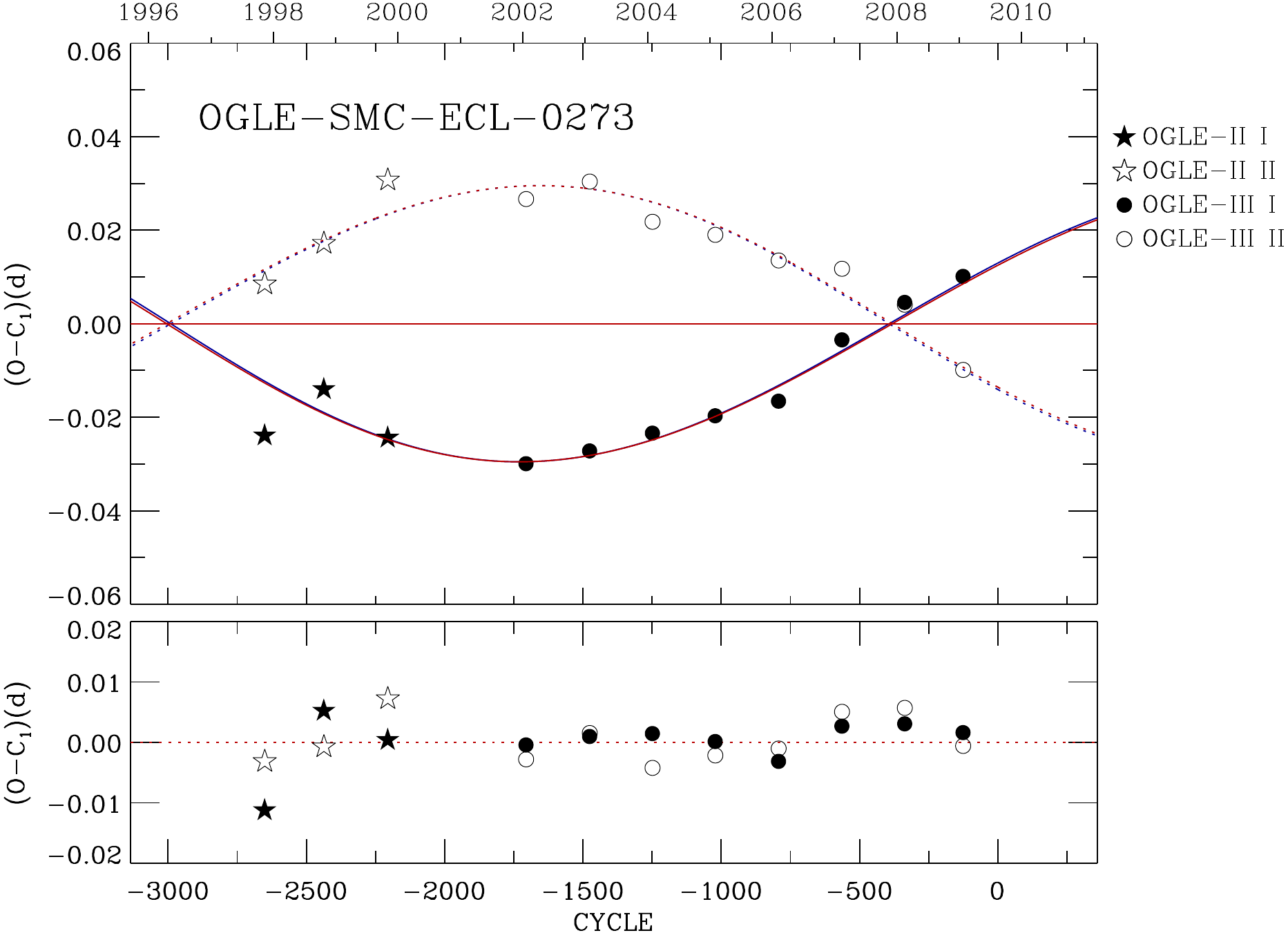} & \includegraphics[width=0.3\columnwidth,height=4.5cm]{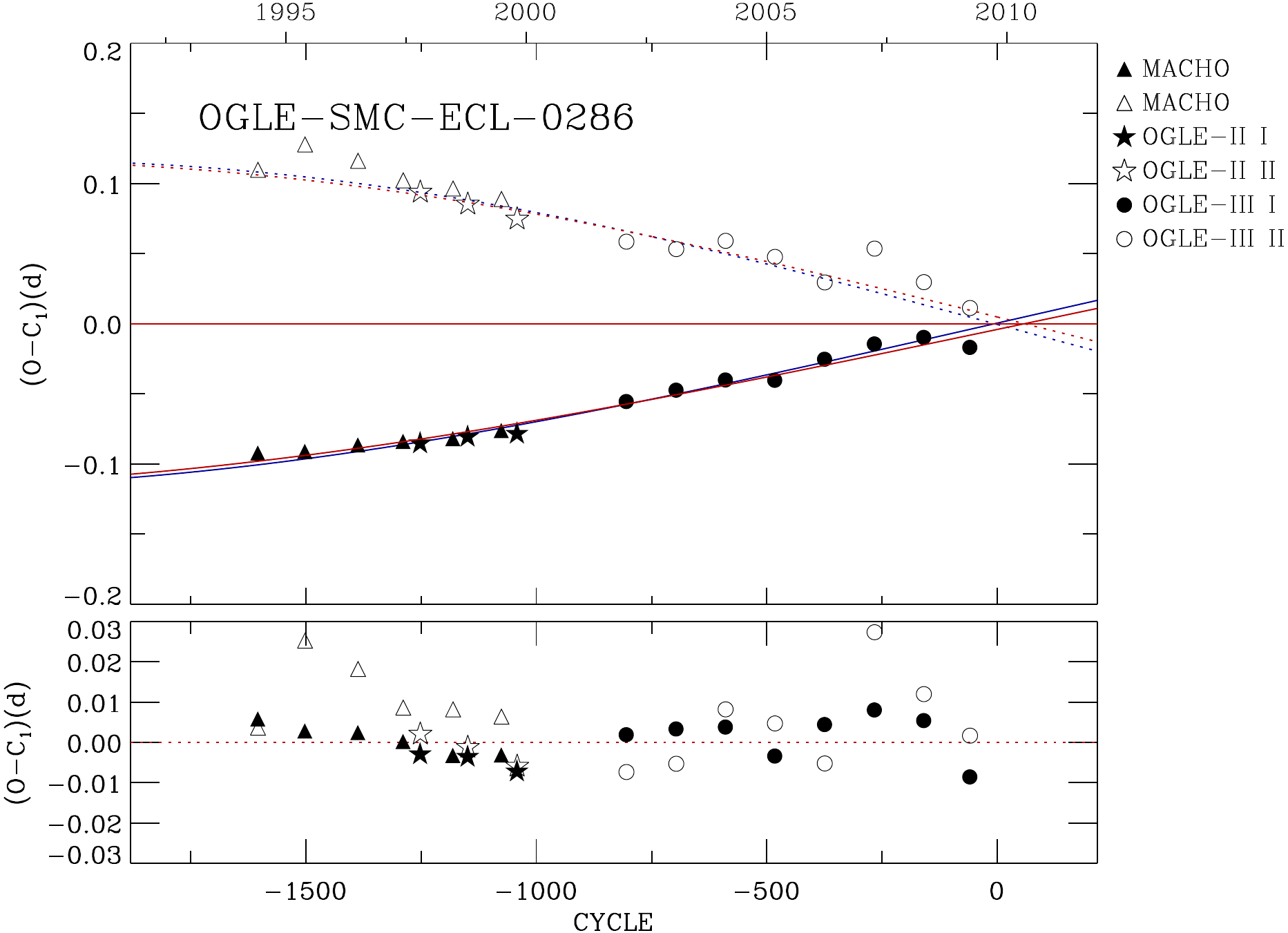}  \\
\includegraphics[width=0.3\columnwidth,height=4.5cm]{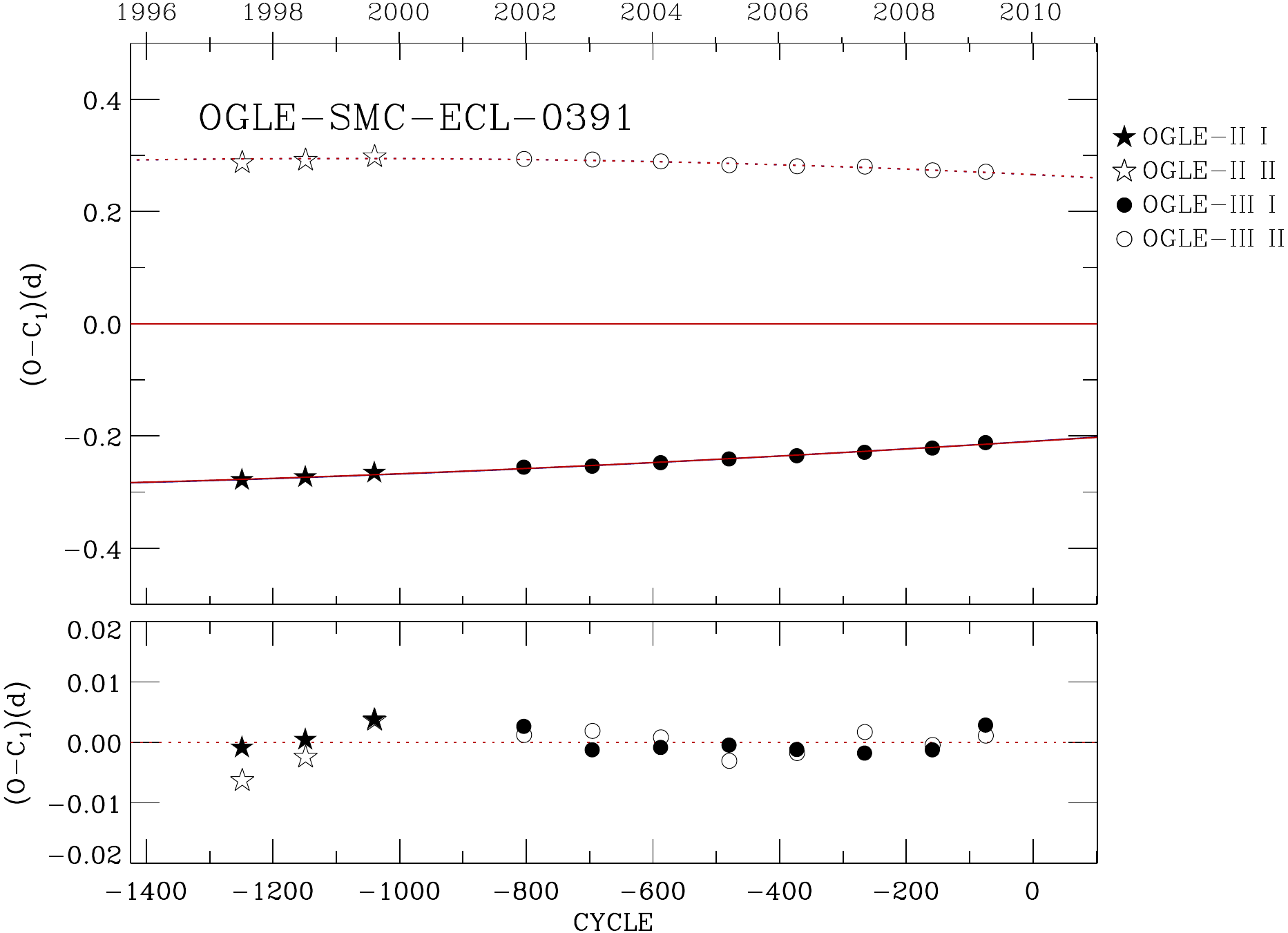} & \includegraphics[width=0.3\columnwidth,height=4.5cm]{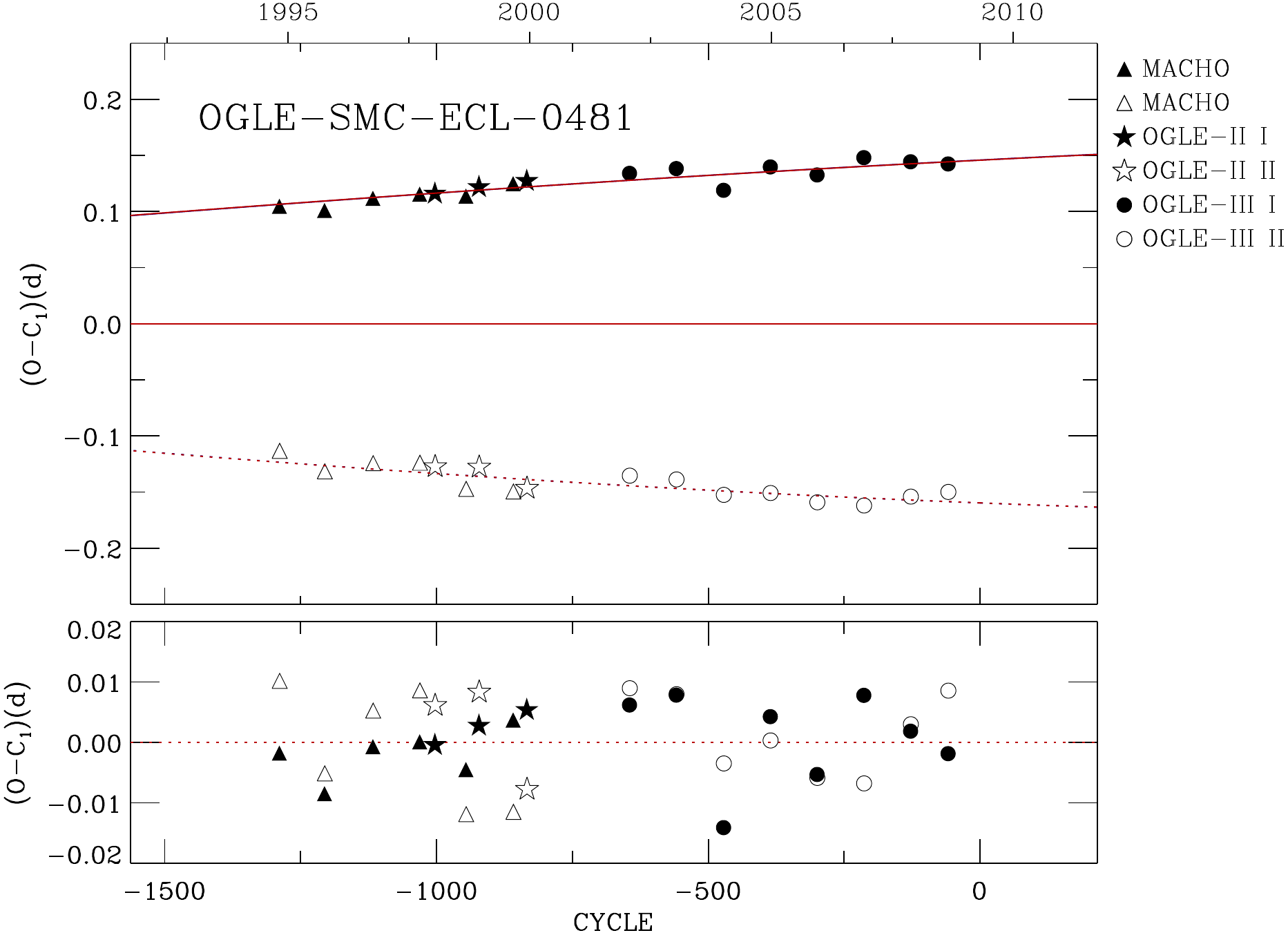} & \includegraphics[width=0.3\columnwidth,height=4.5cm]{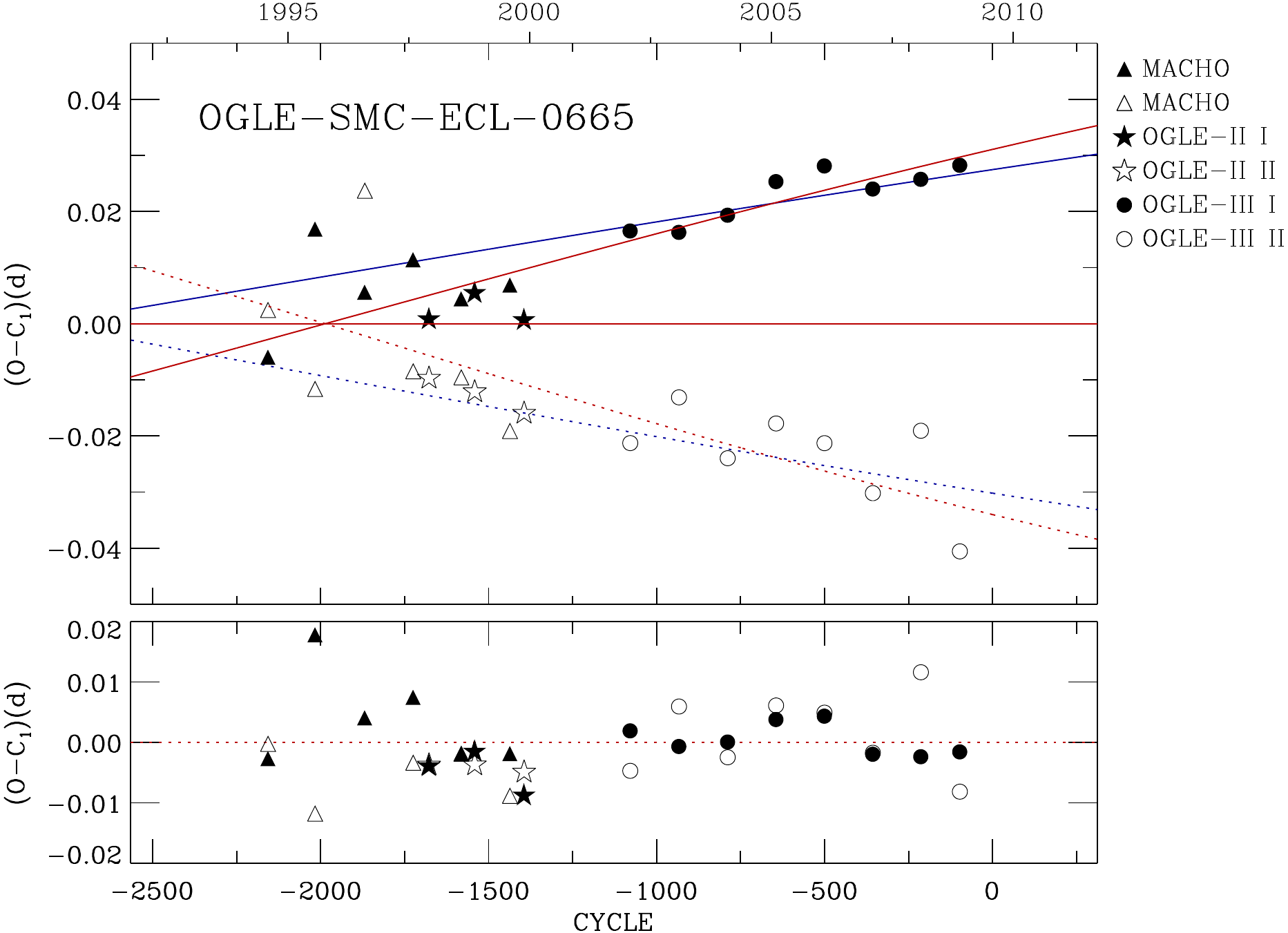}  \\
\includegraphics[width=0.3\columnwidth,height=4.5cm]{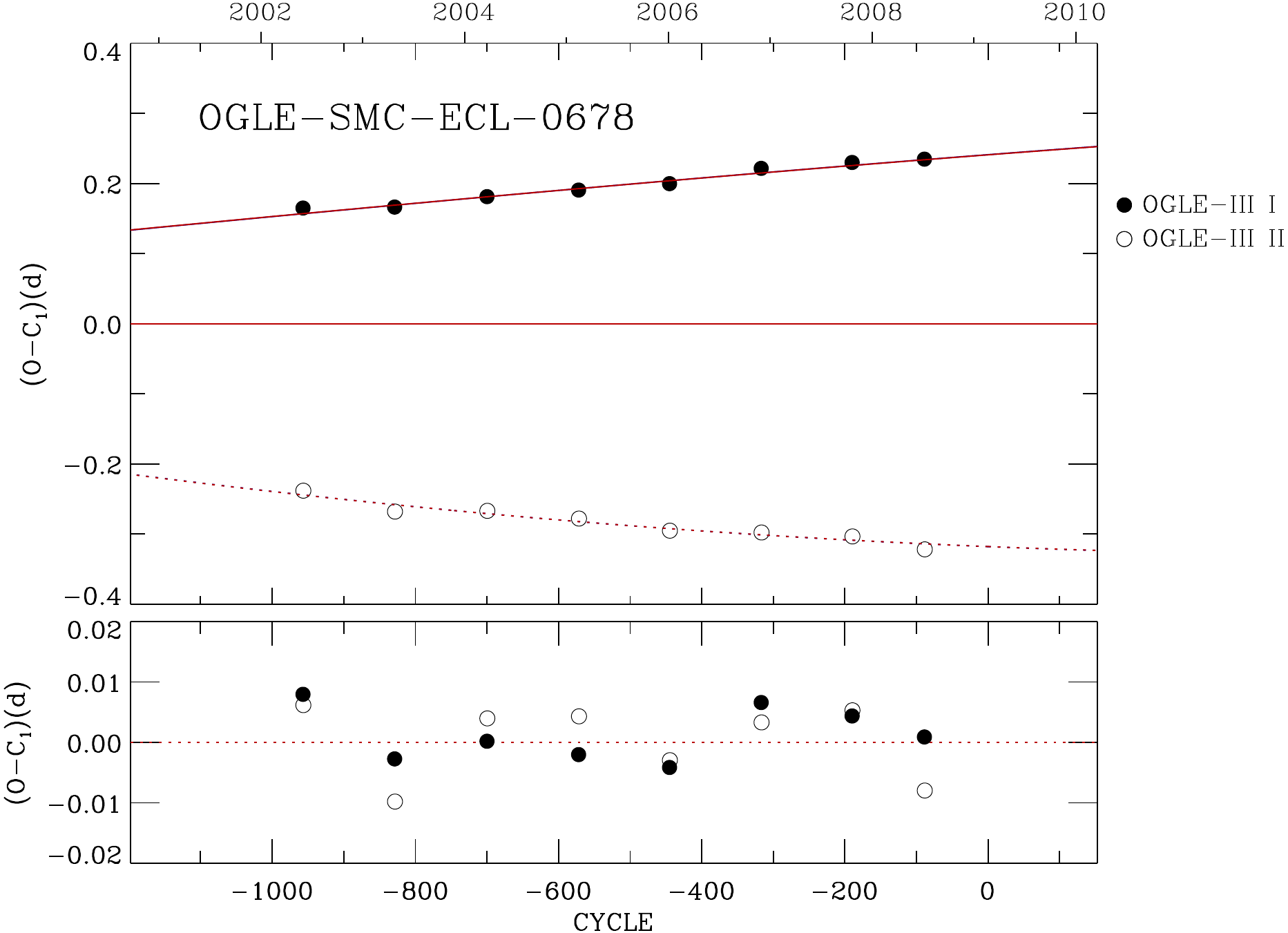} & \includegraphics[width=0.3\columnwidth,height=4.5cm]{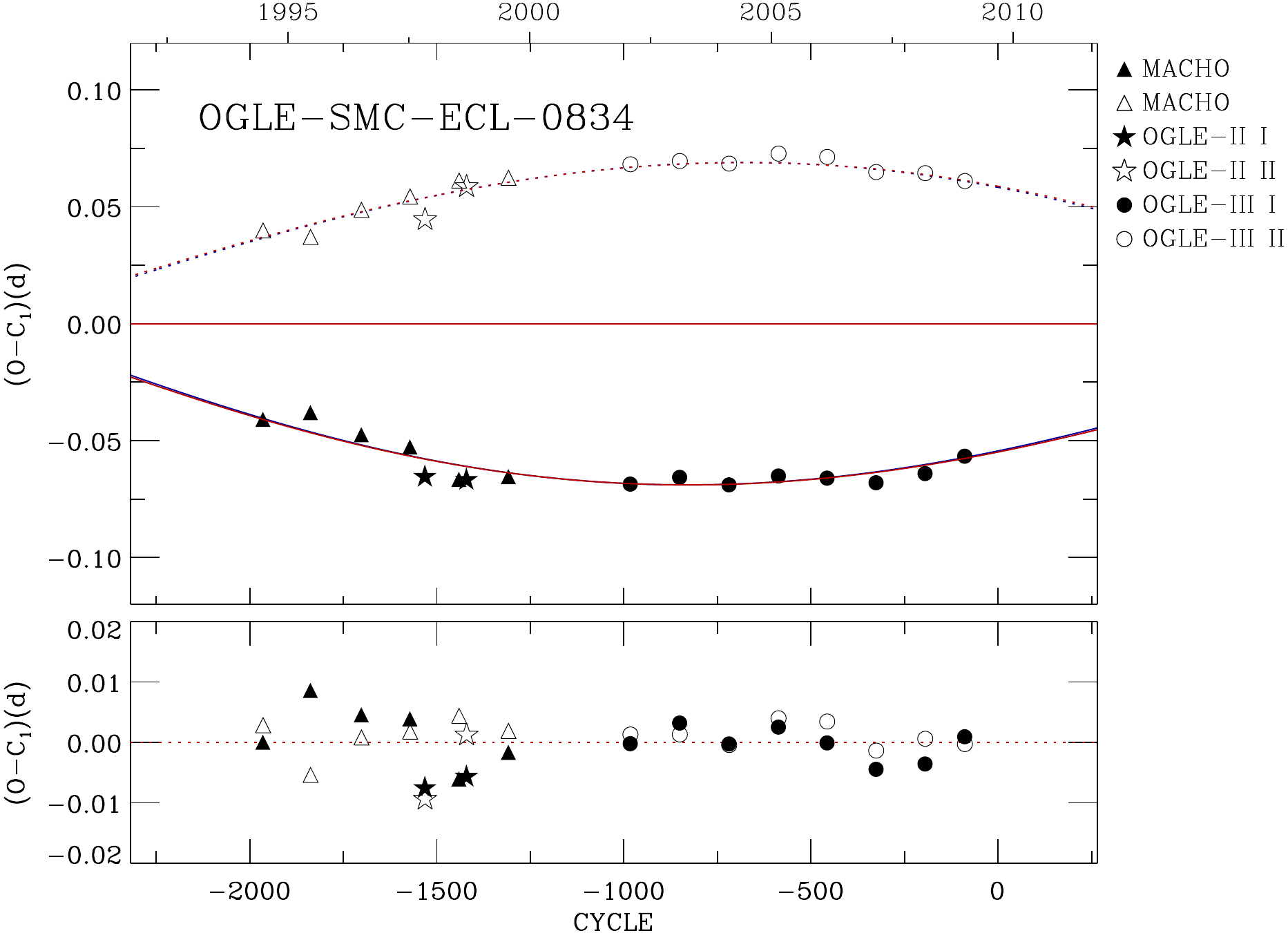} & \includegraphics[width=0.3\columnwidth,height=4.5cm]{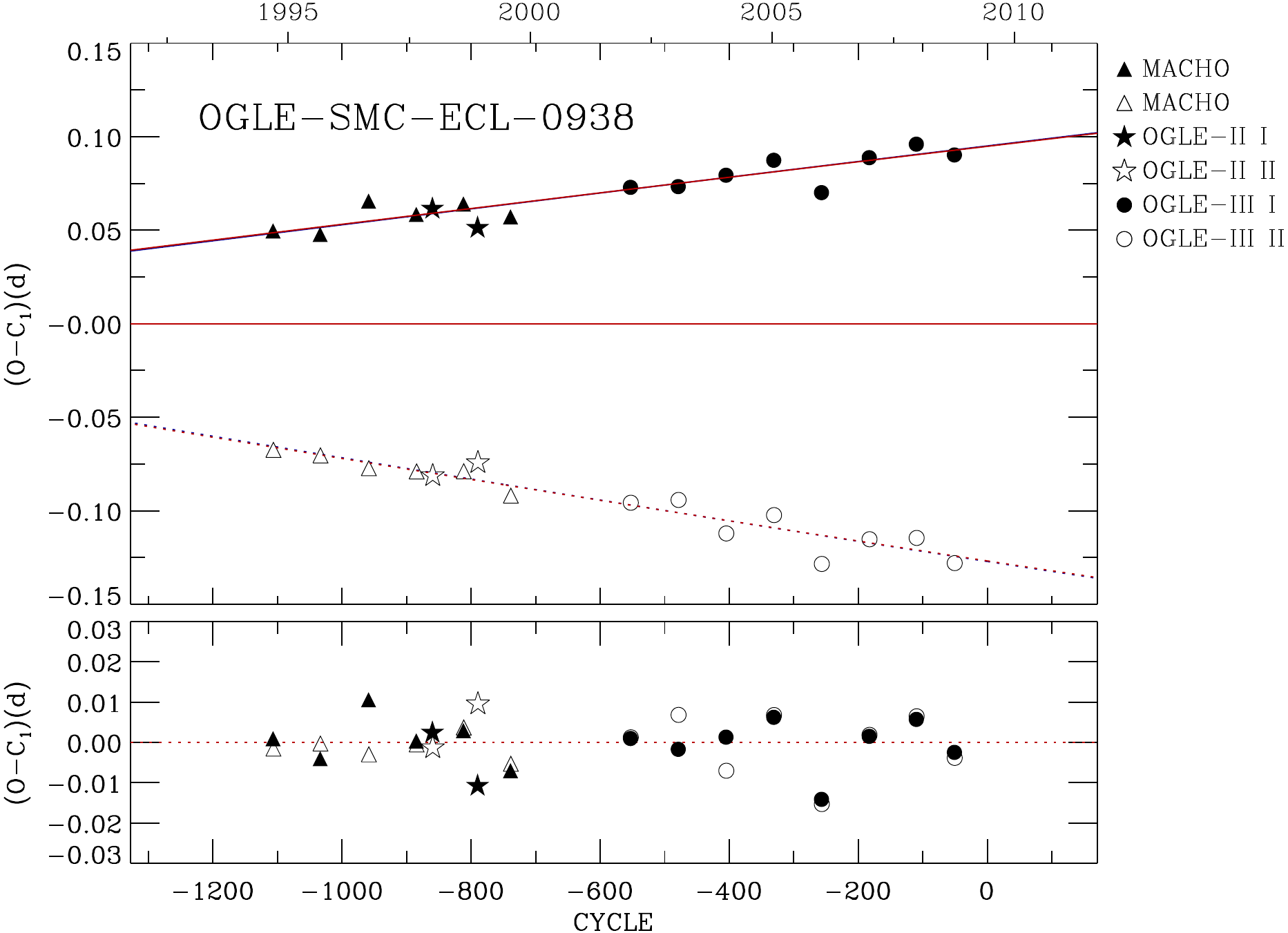}  \\
\end{tabular}
\caption{Eclipse timing diagrams of 9 binary systems. Filled and open symbols represent the individual primary and secondary minima, respectively. In the upper panel, the solid and dashed curves represent the theoretical primary and secondary eclipses of the ephemeris-curve equation, respectively. The red and blue continuous lines represent the eclipse timing and light curve analyses, respectively.
The lower panels display the residuals from the complete ephemeris.}
\end{center}
\end{figure}

\clearpage

\begin{figure}
\begin{center}
\begin{tabular}{cc}
\includegraphics[width=0.4\columnwidth,height=5cm]{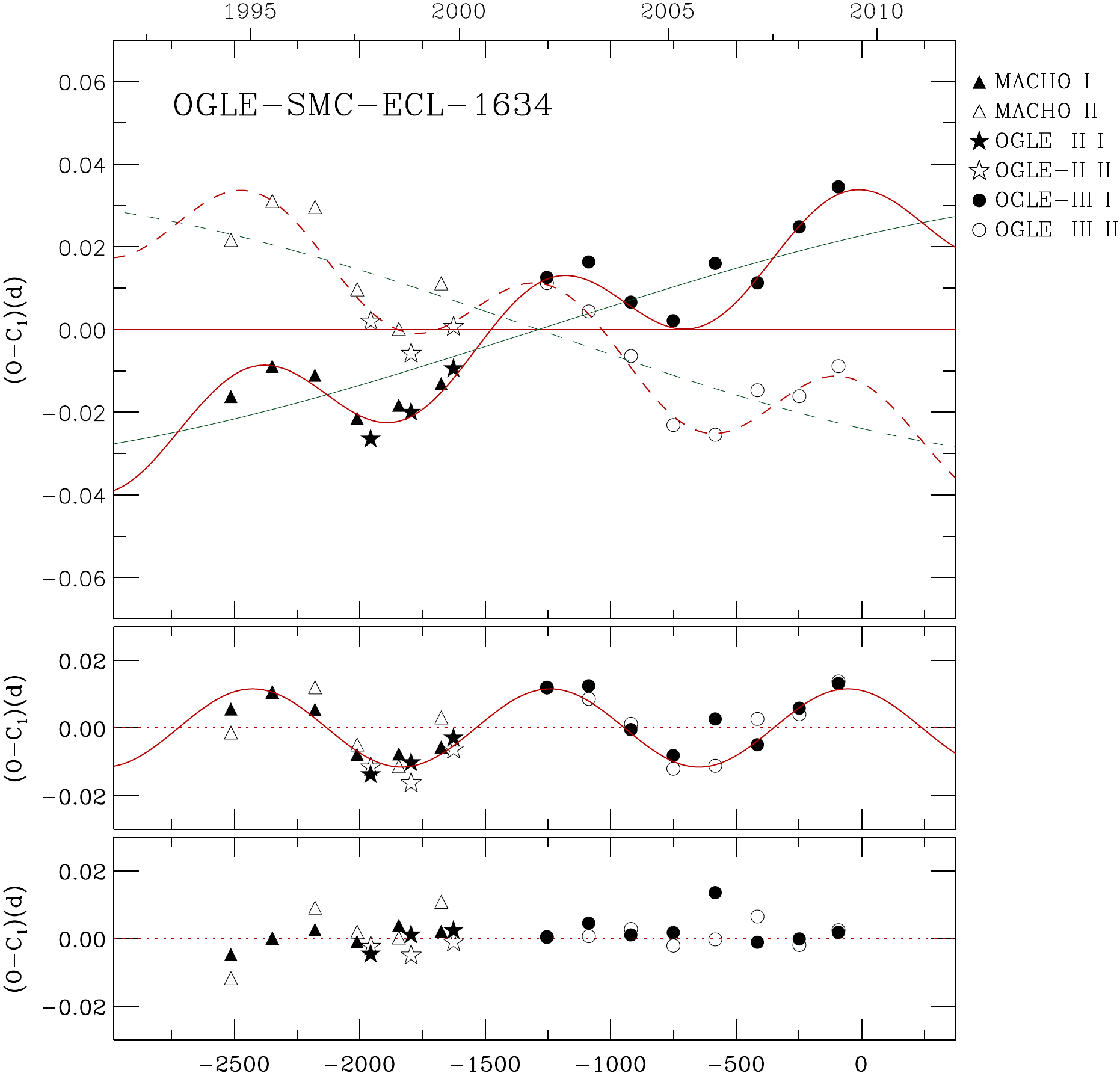} & \includegraphics[width=0.4\columnwidth,height=5cm]{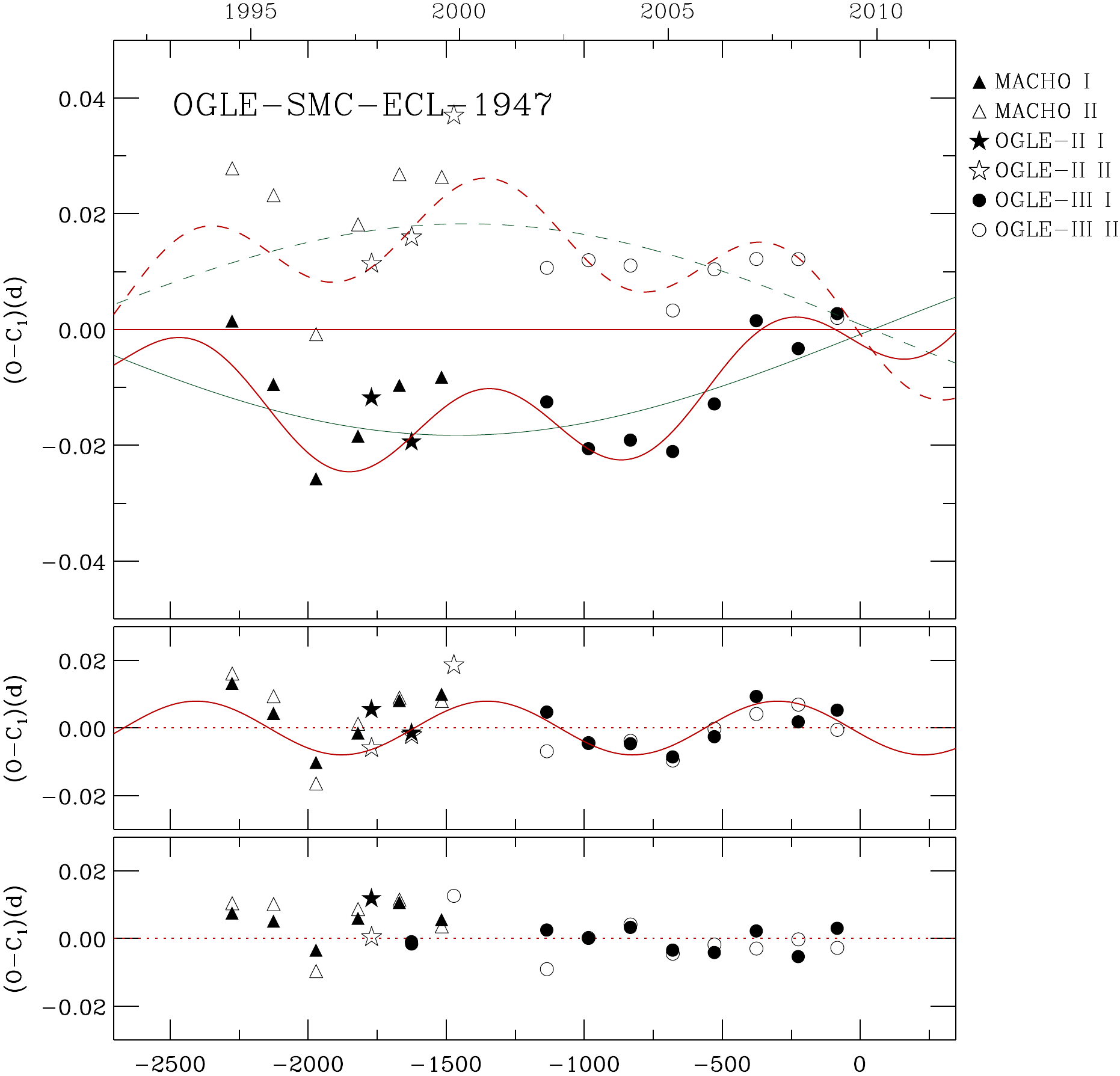} \\ \includegraphics[width=0.4\columnwidth,height=5cm]{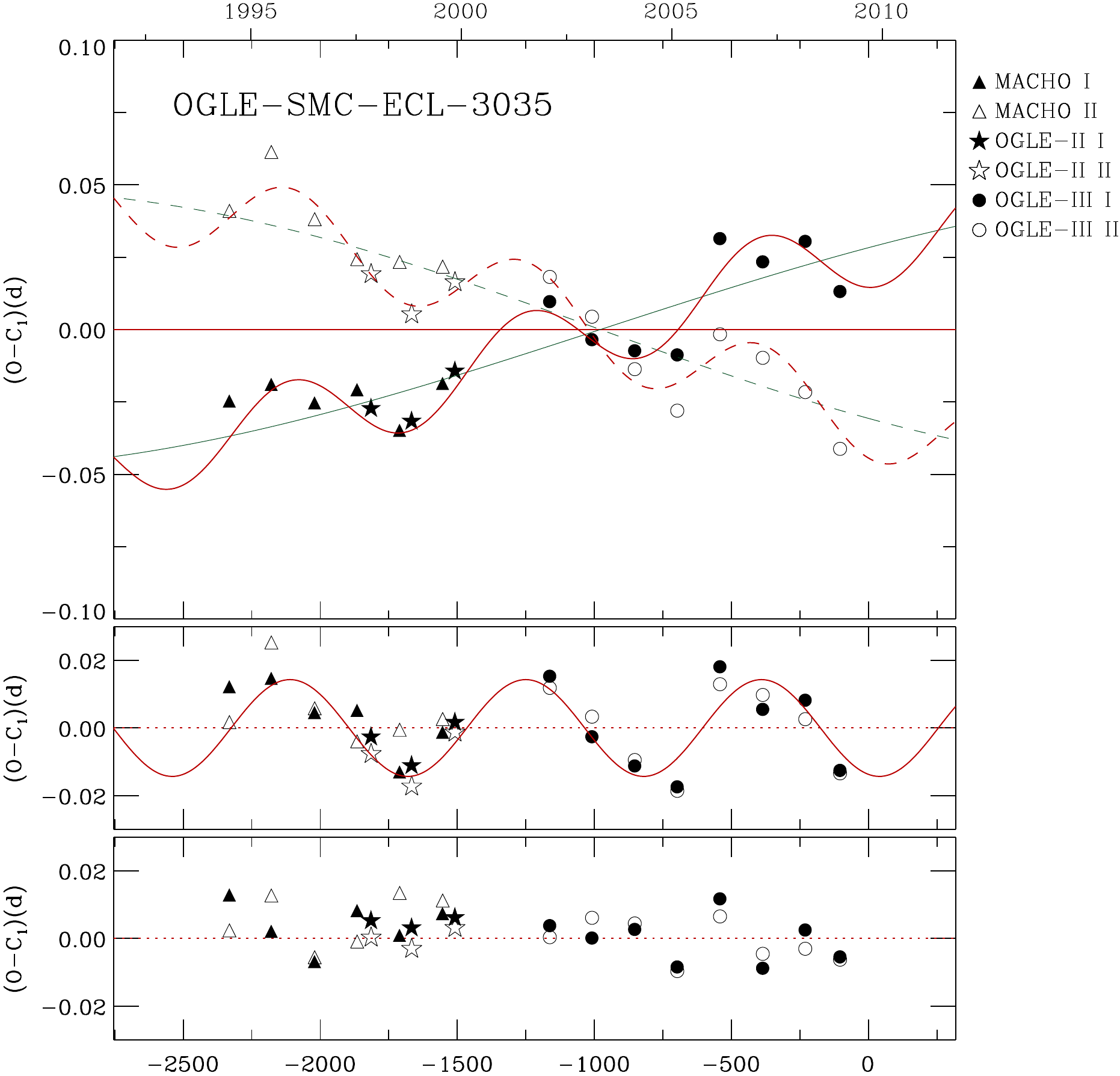} &
\includegraphics[width=0.4\columnwidth,height=5cm]{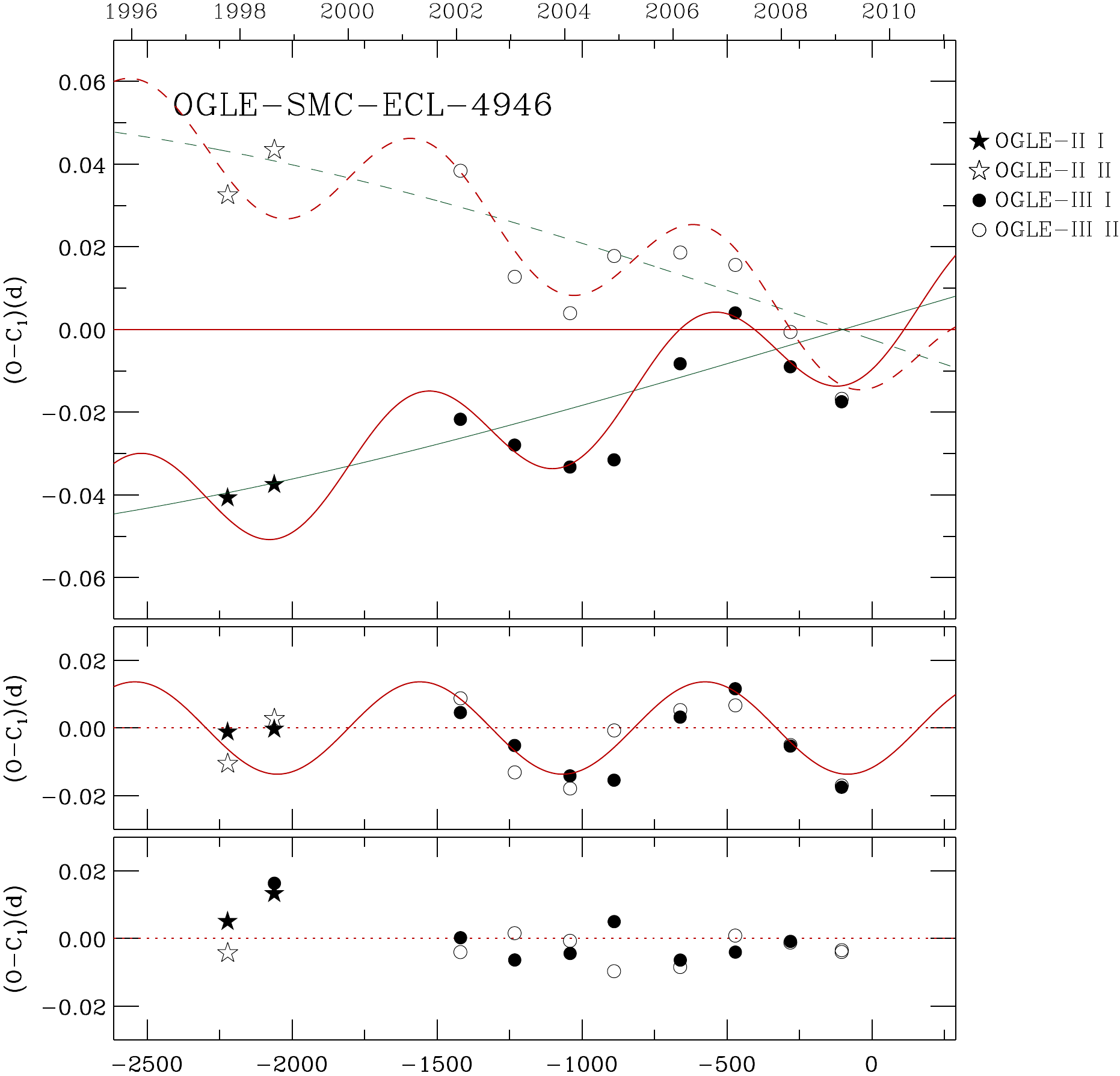}  \\
\end{tabular}
\caption{In the top panels the $O-C$ diagrams of 4 EBs are constructed with the combination effect of the apsidal motion and the additional oscillation for each system. The middle panels display the third-body orbit assuming a circular orbit and the bottom panels represent the residuals after subtracting the two fits.}
\end{center}
\end{figure}


\begin{figure}
\begin{center}
\begin{tabular}{c}
\includegraphics[width=0.7\columnwidth]{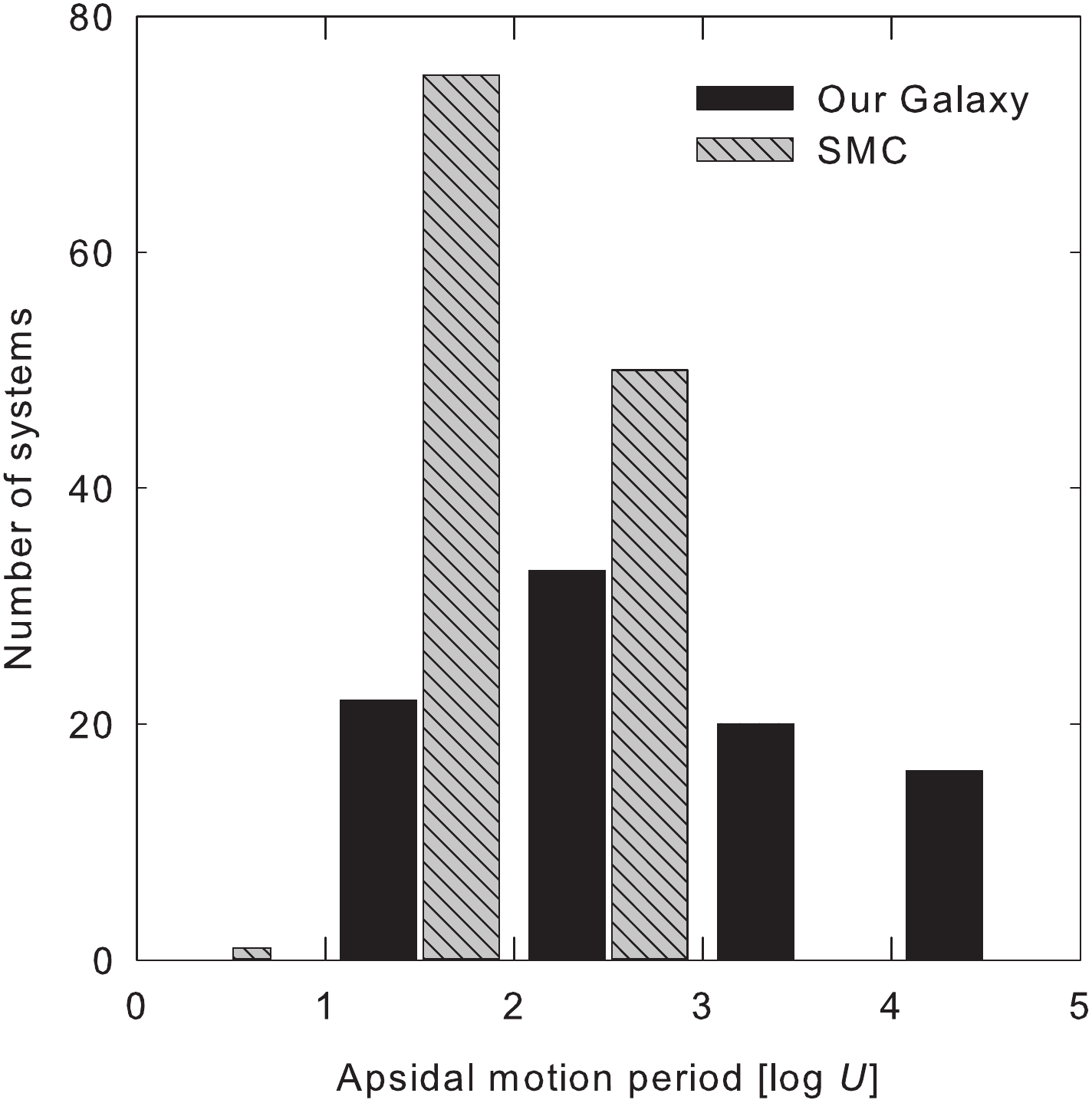}  \\
\end{tabular}
\caption{Histogram of apsidal motion periods of binary systems in our Galaxy and SMC. The black bar represents the 88 detached binaries of our Galaxy in Table E1. The grey bars represent the 126 binary systems taken from Zasche et al. (2014), Hong et al. (2015), and this paper.}
\end{center}
\vspace*{0pt}
\end{figure}

\clearpage

\begin{figure}
\begin{center}
\begin{tabular}{c}
\includegraphics[width=0.7\columnwidth]{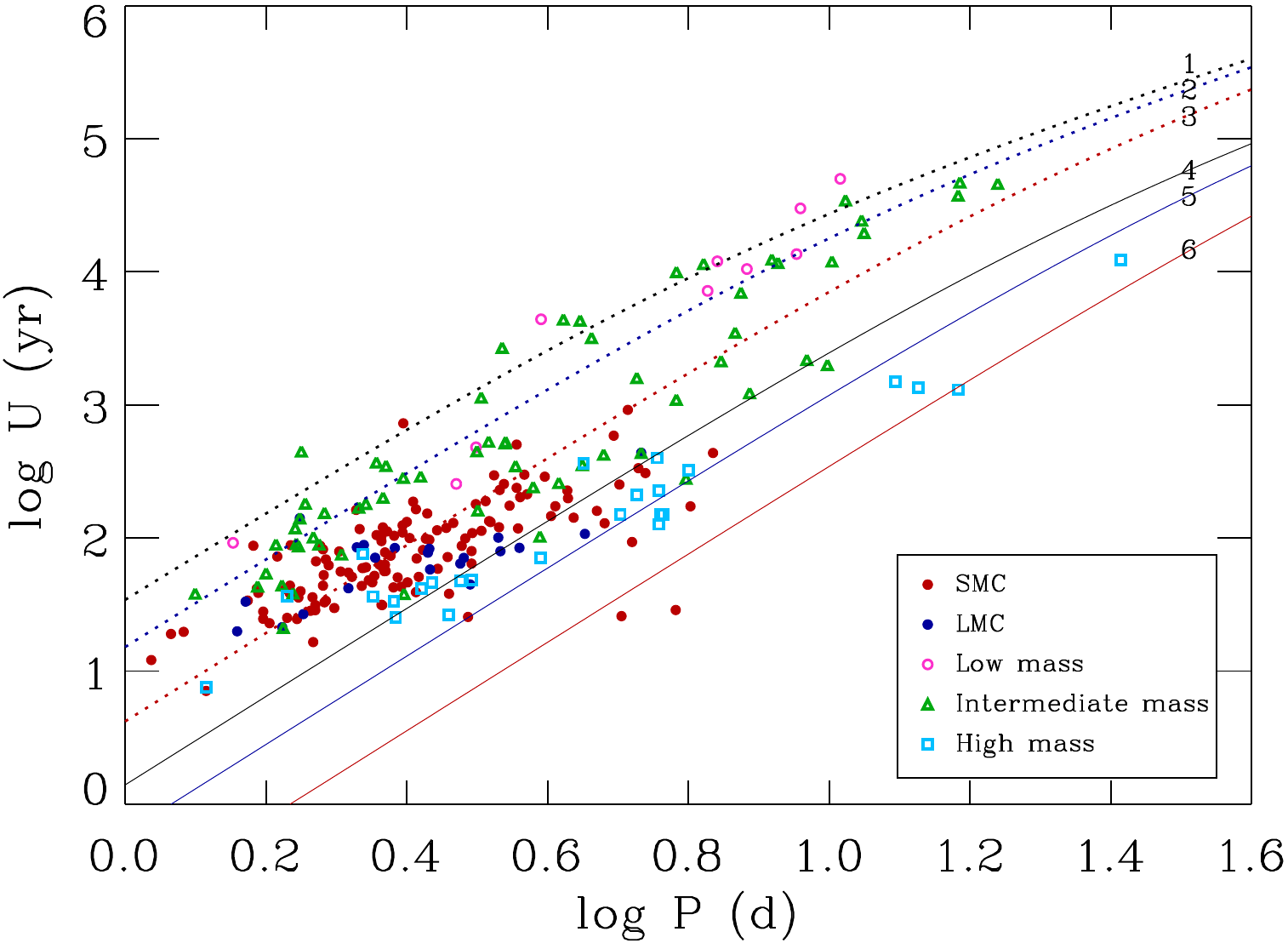}  \\
\end{tabular}
\caption{The orbital period $-$ apsidal period diagram. The red circles display the 126 EBs in the SMC of Zasche et al. (2014), Hong et al. (2015), and this paper. The blue circles represents the 21 EBs  in the LMC taken from Zasche \& Wolf (2013), Hong et al. (2014), and Zasche et al. (2015). The pink circles, green triangles, and cyan squares denote the low ($M < 1.4$ $M_{\odot}$), intermediate (1.4 $M_{\odot} < M < 8 M_{\odot}$), and high mass stars ($M >$ 8 $M_{\odot}$), respectively, of 88 detached binaries in our Galaxy of Table E1. The dotted lines labeled 1, 2, and 3 correspond to $M_1=M_2=1.4$ $M_{\odot}$, and the orbital eccentricities ($e$) of 0.001, 0.3, and 0.5, respectively. The solid lines labeled 4, 5, and 6 correspond to $M_1=M_2=8.0$ $M_{\odot}$, and the values of $e= 0.001$, 0.3, and 0.5, respectively. }
\end{center}
\vspace*{0pt}
\end{figure}

\clearpage

\begin{figure*}
\begin{center}
\begin{tabular}{c}
\includegraphics[width=0.7\columnwidth]{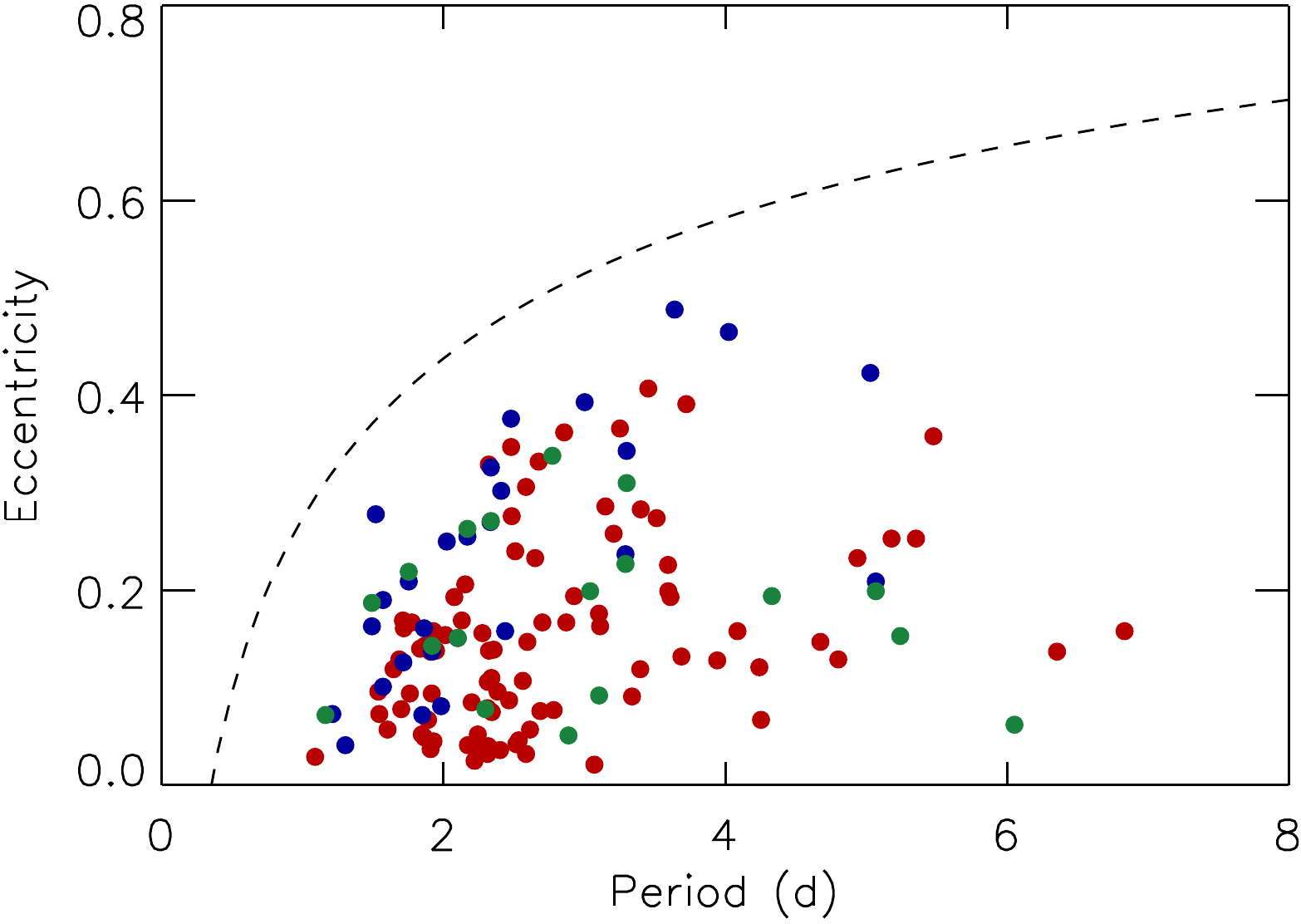}  \\
\end{tabular}
\caption{The orbital period $-$ eccentricity diagram for the SMC EBs. The red circles represent the 90 EBs found in this paper. 
The blue and green circles display the 27 EBs of Hong et al. (2015) and 18 EBs of Zasche et al. (2014), respectively.
 The dashed curve is taken from the equation of $f(P)=E-A\times {\rm exp}^{-(P\times B)^C}$, where  $E=0.98, A=3.25, B=6.3$ and $C=0.23$. }
\end{center}
\vspace*{0pt}
\end{figure*}

\clearpage
\begin{deluxetable}{lccccccc}
\tabletypesize{\scriptsize}
\tablewidth{-20pt}
\tablecaption{Basic Parameters of 90 Binary Systems.}
\tablehead{
\colhead{OGLE-III}  &\colhead{OGLE-II}& \colhead{MACHO}   &\colhead{R.A.}       &\colhead{Decl.}           &\colhead{$I$}     & \colhead{$V$}      &\colhead{$V-I$}  \\
\colhead{(ID)}      &\colhead{(ID)}   & \colhead{(ID)}    &\colhead{(J2000)}  &     \colhead{(J2000)}      &\colhead{(mag)}   & \colhead{(mag)}    &\colhead{(mag)}        
}
\startdata
OGLE-SMC-ECL-0237  &                  &                   &    00:35:31.21  & $-72$:58:03.6 &  17.854(26) &  17.744(11) &  $ -0.110$(19)  \\
OGLE-SMC-ECL-0273  &  SMC-SC1-9354    &                   &    00:36:18.09  & $-73$:33:15.5 &  17.237(16) &  17.049(7)  &  $ -0.188$(12)  \\
OGLE-SMC-ECL-0286  &  SMC-SC1-25589   &   213.15110.69    &    00:36:39.63  & $-73$:01:59.0 &  17.328(18) &  17.328(8)  &  $  0.000$(13)  \\
OGLE-SMC-ECL-0391  &  SMC-SC1-97290   &                   &    00:39:25.29  & $-73$:38:36.1 &  16.623(11) &  16.490(5)  &  $ -0.133$(8)   \\
OGLE-SMC-ECL-0481  &  SMC-SC2-74863   &   213.15396.67    &    00:41:03.31  & $-72$:58:45.9 &  17.286(17) &  17.236(9)  &  $ -0.050$(13)  \\
OGLE-SMC-ECL-0665  &  SMC-SC3-80396   &   213.15563.267   &    00:43:29.09  & $-73$:15:43.1 &  17.873(24) &  17.966(14) &  $  0.093$(19)  \\
OGLE-SMC-ECL-0678  &                  &                   &    00:43:38.97  & $-73$:42:57.2 &  17.789(27) &  17.683(10) &  $ -0.106$(19)  \\
OGLE-SMC-ECL-0834  &  SMC-SC3-189164  &   213.15674.20    &    00:45:10.19  & $-73$:25:47.1 &  16.716(11) &  16.672(7)  &  $ -0.044$(9)   \\
\enddata
\end{deluxetable}

\clearpage
\renewcommand{\tabcolsep}{0.6mm}
\begin{deluxetable}{lccccccccccccc}
\tabletypesize{\tiny} 
\tablewidth{-30pt}
\tablecaption{Orbital parameters from the light curve analyses}
\tablehead{
\colhead{Object ID} &\colhead{$T_2/T_1$} &\colhead{$e$}  &\colhead{$\omega$ }&\colhead{$\dot\omega$}&\colhead{$q$ $(M_2/M_1)$ }   &\colhead{$i$ }   &\colhead{$\Omega_{1}$ } &  \colhead{$\Omega_{2}$} &\colhead{$l^b_{1,I}$} &\colhead{$l^b_{2,I}$}  & \colhead{$l^b_{3,I}$}  & \colhead{$r_{1}$$\rm ^c$ } &\colhead{$r_{2}$$\rm ^c$  }    \\
\colhead{(OGLE-)}   &                      &               &  (deg)            &(deg year$^{-1})$     &                 & (deg)           &\colhead{ }             & \colhead{ }             &                      &                       &                        & \colhead{(volume)}         &\colhead{(volume)}     
}                                
\startdata                
SMC-ECL-0237   &    0.935(20)  &  0.301(4)   &    70(1)  &    2.16(21)    & 1.19(14)  &   84.4(3)   &    9.55(27)   &    7.13(60) &  0.299(9)   &  0.701 &   $-$           &  0.1276(56)   &  0.2071(56)    \\  
SMC-ECL-0273   &    0.974(15)  &  0.056(3)   &   117(4)  &   15.81(1.55)  & 0.84(10)  &   77.5(2)   &    5.06(16)   &    5.91(51) &  0.660(23)  &  0.340 &   $-$           &  0.2417(113)  &  0.1772(113)   \\  
SMC-ECL-0286   &    0.978(7)   &  0.107(5)   &    88(2)  &    4.30(42)    & 0.65(4)   &   82.8(4)   &    4.73(7)    &    6.87(32) &  0.831(4)   &  0.169 &   $-$           &  0.2517(51)   &  0.1159(51)    \\  
SMC-ECL-0391   &    0.949(7)   &  0.269(1)   &    36(1)  &    2.40(21)    & 0.74(6)   &   87.7(8)   &    6.04(11)   &    7.18(41) &  0.469(21)  &  0.185 &   0.346(0.022)  &  0.1998(54)   &  0.1311(54)    \\  
SMC-ECL-0481   &    1.087(14)  &  0.125(5)   &   152(5)  &    1.40(22)    & 1.12(8)   &   78.6(2)   &    6.91(14)   &    7.20(43) &  0.447(20)  &  0.553 &   $-$           &  0.1784(54)   &  0.1843(54)    \\  
SMC-ECL-0665   &    1.048(13)  &  0.071(14)  &   117(7)  &    1.53(43)    & 0.91(10)  &   81.9(2)   &    5.99(19)   &    6.76(57) &  0.585(28)  &  0.415 &   $-$           &  0.2003(91)   &  0.1609(91)    \\  
SMC-ECL-0678   &    0.880(17)  &  0.357(1)   &   149(1)  &    3.02(24)    & 0.71(5)   &   85.4(4)   &    7.34(11)   &    8.47(43) &  0.744(17)  &  0.256 &   $-$           &  0.1608(35)   &  0.1056(35)    \\  
SMC-ECL-0834   &    0.985(3)   &  0.076(1)   &    32(3)  &    6.10(21)    & 0.91(3)   &   78.6(1)   &    5.21(05)   &    5.57(13) &  0.579(8)   &  0.421 &   $-$           &  0.2386(32)   &  0.2063(32)    \\
\enddata                                                                                                                                                                                                                                                                                           
\begin{flushleft}
Notes$-$Listed parameters: temperature ratio, eccentricity, longitude of periastron, rate of periastron advance, mass ratio, inclination, dimensionless surface potentials, $I$ band luminosity ratios, and mean volume radii. All quoted uncertainties of the photometric solutions were calculated from the Differential Corrections subroutine of the Wilson-Devinney Code. The luminosity ratios of the other band ($V_MR_MVI$) are listed in Table C.
\newline
$^{a}$Fixed parameter.
\newline
$^{b}$$l_{x}/(l_{1}+l_{2}+l_{3})$.
\newline
$^{c}$Mean volume radius.
\end{flushleft}
\end{deluxetable}



\clearpage
\renewcommand{\arraystretch}{1}
\renewcommand{\tabcolsep}{1mm}
\begin{deluxetable}{lccccc}
\tablewidth{0pt}
\tablecaption{List of eclipse timings for 90 EBs}
\tablehead{
\colhead{Object ID}    &\colhead{JD Hel.}     &  \colhead{Error} &\colhead{Cycle} & \colhead{Min}  &\colhead{Source} \\
\colhead{(OGLE-)}     &\colhead{(+2400000)} & \colhead{(day)}    &\colhead{ }       &                &       
}       
\startdata       
  SMC-ECL-0237  &   52274.6757  &   0.0026  &  $ -1055.0 $  & I  & OGLE-III \\ 
                &   52636.6385  &   0.0024  &  $  -915.0 $  & I  & OGLE-III \\ 
                &   52998.6088  &   0.0021  &  $  -775.0 $  & I  & OGLE-III \\ 
                &   53363.1647  &   0.0037  &  $  -634.0 $  & I  & OGLE-III \\ 
                &   53730.3011  &   0.0032  &  $  -492.0 $  & I  & OGLE-III \\ 
                &   54097.4468  &   0.0041  &  $  -350.0 $  & I  & OGLE-III \\ 
                &   54459.4155  &   0.0028  &  $  -210.0 $  & I  & OGLE-III \\ 
                &   54751.5771  &   0.0025  &  $   -97.0 $  & I  & OGLE-III \\ 
                &   52276.2634  &   0.0026  &  $ -1054.5 $  & II & OGLE-III \\ 
                &   52638.2202  &   0.0024  &  $  -914.5 $  & II & OGLE-III \\ 
                &   53000.1745  &   0.0021  &  $  -774.5 $  & II & OGLE-III \\ 
                &   53364.7172  &   0.0037  &  $  -633.5 $  & II & OGLE-III \\ 
                &   53731.8340  &   0.0032  &  $  -491.5 $  & II & OGLE-III \\ 
                &   54098.9565  &   0.0041  &  $  -349.5 $  & II & OGLE-III \\ 
                &   54460.9040  &   0.0028  &  $  -209.5 $  & II & OGLE-III \\ 
                &   54753.0555  &   0.0025  &  $   -96.5 $  & II & OGLE-III \\ 
\enddata            
\begin{flushleft}
(This table is available in its entirety in machine-readable and Virtural Observatory (VO) forms.)
\end{flushleft}
\end{deluxetable}


\clearpage
\renewcommand{\arraystretch}{1}
\renewcommand{\tabcolsep}{1mm}
\begin{deluxetable}{lcccccccc}
\tabletypesize{\small}
\tablewidth{0pt}
\tablecaption{Apsidal motion elements of 90 EBs}
\tablehead{
\colhead{Object ID}  &\colhead{$T_{0}$}  &\colhead{$P_{s}$} &\colhead{$P_{a}$}  & \colhead{$\omega_{0}$}  &\colhead{$\dot\omega$} &\colhead{ $U^{a}$}  &\colhead{ $U^{b}$} &\colhead{ $U^{c}$}   \\
\colhead{(OGLE-)}     &\colhead{(+2450000)} & \colhead{(day)}    &\colhead{(day) }   &  (deg)                  &  (deg year$^{-1}$)      &  (year)   &  (year)    &  (year)  
}       
\startdata    
SMC-ECL-0237  &   5002.4367(14)    &  2.585443(5)    &    2.585557(5)     &     71(1)   &    2.26(15)     &    160(10)  &    167(15)  &      162(8)  \\
SMC-ECL-0273  &   5001.5390(10)    &  1.602491(6)    &    1.602797(6)     &    116(4)   &   15.63(56)     &     23(1)   &      23(2)  &       23(1)  \\
SMC-ECL-0286  &   5003.4038(10)    &  3.396074(7)    &    3.396422(7)     &     88(2)   &    3.97(16)     &     91(3)   &      84(7)  &       90(3)  \\
SMC-ECL-0391  &   5001.2617(14)    &  3.399923(13)   &    3.400132(13)    &     36(6)   &    2.37(32)     &    152(18)  &    150(12)  &     150(10)  \\
SMC-ECL-0481  &   5003.9739(6)     &  4.241589(6)    &    4.241780(6)     &    153(4)   &    1.39(8)      &    258(15)  &    257(35)  &     258(14)  \\
SMC-ECL-0665  &   5002.4193(6)     &  2.533936(6)    &    2.534058(6)     &    124(2)   &    2.49(24)     &    145(13)  &    235(52)  &     150(13)  \\
SMC-ECL-0678  &   5002.6053(13)    &  2.856128(5)    &    2.856315(5)     &    149(1)   &    3.02(16)     &    119(6)   &     119(9)  &      119(5)  \\
SMC-ECL-0834  &   5001.9925(11)    &  2.781443(12)   &    2.781797(12)    &     34(1)   &    6.01(39)     &     60(4)   &      59(2)  &       59(2)  \\
\enddata
\begin{flushleft}
Notes$-$Listed parameters: zero epoch, sidereal period, anomalistic period, longitude of periastron, rate of periastron advance, and periods of apsidal motions.  
\tablenotetext{a}{Determined from the eclipse timing analysis.}
\tablenotetext{b}{Determined from the light curve analysis.}
\tablenotetext{c}{The weighted mean from $U^a$ and $U^b$.  }
\end{flushleft}
\end{deluxetable}


\clearpage
\renewcommand{\arraystretch}{1}
\renewcommand{\tabcolsep}{1mm}
\begin{deluxetable}{lccccc}
\tablewidth{0pt}
\tablecaption{The third-body parameters for 4 EBs assuming circular orbits.}
\tablehead{\colhead{Parameter }   &\colhead{Unit}  &\colhead{SMC-ECL-1634}     &  \colhead{SMC-ECL-1947} &\colhead{SMC-ECL-3035}  & \colhead{SMC-ECL-4946}  
}       
\startdata    
$K$               &    days            &     0.012(3)   &   0.008(5)    &    0.014(5)    &     0.014(4)      \\ 
$\Omega$          &    deg $P^{-1}$    &     0.30(1)    &    0.34(2)    &    0.42(2)     &     0.37(2)       \\   
$\Omega_0$        &    deg             &     107(16)    &    192(29)    &    253(18)     &     301(18)         \\      
$P_3$             &    yr              &     7.1(7)     &   6.9(1.2)    &     5.5(5)     &      5.2(4)         \\      
$a_{12}$sin$i_3$  &    AU              &     2.0(5)     &    1.4(9)     &     2.5(8)     &      2.4(7)       \\      
f($M_3$)          &    $M_{\odot}$     &     0.16(4)    &    0.05(4)    &     0.50(17)   &      0.50(15)     \\      
\enddata
\end{deluxetable}

\label{lastpage}
\end{document}